\documentclass[prl,aps,twocolumn,superscriptaddress,nofootinbib]{revtex4}
\usepackage{graphicx} % Include figure files
\usepackage{dcolumn}  % Align table columns on decimal point
\usepackage{bm}       % bold math
\usepackage{amsmath}
\usepackage{amssymb}
\usepackage{epsfig}
\usepackage{color}
\usepackage{bbold}

\begin{document}
\title{Finite-Size Effects on Periodic Arrays of Nanostructures}
\author{Lauren Zundel}
\affiliation{Department of Physics and Astronomy, University of New Mexico, Albuquerque, New Mexico 87131, United States}
\author{Alejandro Manjavacas}
\email[Corresponding author: ]{manjavacas@unm.edu}
\affiliation{Department of Physics and Astronomy, University of New Mexico, Albuquerque, New Mexico 87131, United States}

\date{\today}

\begin{abstract}
Arrays of nanostructures have emerged as exceptional tools for the manipulation and control of light. Oftentimes, despite the fact that real implementations of nanostructure arrays must be finite, these systems are modeled as perfectly periodic, and therefore infinite. Here, we investigate the legitimacy of this approximation by studying the evolution of the optical response of finite arrays of nanostructures as their number of elements is increased. We find that the number of elements necessary to reach the infinite array limit is determined by the strength of the coupling between them, and that, even when that limit is reached, the individual responses of the elements may still display significant variations.  In addition, we show that, when retardation is negligible, the resonance frequency for the infinite array is always redshifted compared to the single particle. However, in the opposite situation, there could be either a blue- or a redshift. We also study the effects of inhomogeneity in size and position of the elements on the optical response of the array. This work advances the understanding of the behavior of finite and infinite arrays of nanostructures, and therefore provides guidance to design applications that utilize these systems.
\end{abstract}
\maketitle

Metallic nanoparticles, capable of supporting surface plasmon modes, have proven to be ideal tools for manipulating light, 
due to their strong optical responses and subwavelength field confinement \cite{M07}. These exceptional properties are being exploited in a wide variety of applications, including ultrasensitive biosensing \cite{XBK99,AHL08}, solar energy harvesting \cite{CP08,AP10}, photocatalysis \cite{BQ14,BHN15}, nanoscale light emission \cite{AOS12,LLR13,ZDS13,YO15,HRV17}, imaging \cite{ASV06,WWS17}, and nonlinear optics \cite{NBK12,FZP06,MGF15}, to cite a few.  Metallic nanostructures are also used as building blocks for metasurfaces \cite{YC14}, which are ultrathin structures that enable the manipulation of the wavefront of light beams on a subwavelength scale \cite{YGK11}. 

Most of these applications involve the use of ensembles of metallic nanostructures, which are commonly arranged in periodic geometries \cite{WRV18}. This, in addition to providing a response stronger than that of a single nanostructure, can also lead to collective behaviors arising from coherent interactions between the nanostructures \cite{paper090,JRP17,HMV18}. That is the case for so-called lattice resonances, which occur at wavelengths commensurable with the periodicity of the array \cite{AB08,VGG09-2,paper164,HB14,HB16,KK14,ama59,KKB18}, and have particularly strong and narrow spectral features that make them ideal for the previously mentioned applications \cite{KSG08,GVG10,RLV12,SK14_2,SK14,LGV14,COK16}.

Usually, arrays of nanostructures are modeled as though they were perfectly periodic, and, consequently, infinite. By doing so, it is possible to take full advantage of periodicity, and therefore to calculate the response of the whole system by only modeling the unit cell of the array \cite{ama59}. This significantly reduces the computational cost of the calculation as compared with the modeling of each element in a finite array \cite{paper090,ama41}. In reality, however, no array is infinite, and, usually, the size of the arrays that can be created in the laboratory is limited by the employed fabrication method.  This can lead to significant discrepancies between the optical response of the modeled, perfectly periodic, system, and that of the fabricated one, which arise from both the effect of the boundaries, present in finite systems but not infinite ones, as well as from the truncation of the collective behavior caused by the finiteness of the structure \cite{SHV08,FPP10,NBS11,RSB12,M17,MMT17}.

%This can lead to a significant discrepancy between the optical response of the modeled, perfectly periodic, system, and that of the fabricated one, which arises from the effect that the boundaries, present in finite systems, but not for infinite ones, have on the collective response of the array \cite{RSB12}.

Here, we seek to understand when the finite-size effects on arrays of nanoparticles can be neglected, and their response can therefore be modeled assuming they are perfectly periodic. To this end, we use a coupled dipole model to analyze the optical response of finite arrays of metallic nanostructures, made of gold, silver, or graphene, with a varying number of elements. We consider arrays in which the nanostructures are separated by distances smaller than the resonance wavelength, as well as those in which the separation is comparable to it, and therefore can support lattice resonances.  By comparing the response of these systems with that of the corresponding infinite arrays, we find that the number of elements required to reach the infinite array limit is determined by the strength of the interaction between them. Furthermore, we show that, even when the collective response has converged to the infinite array limit, the individual responses of the constituents may still differ greatly from the perfectly periodic case. We also demonstrate that, depending on the role played by retardation, the resonance frequency of the infinite array can be either red- or blueshifted compared to that of a single nanostructure. We analyze, as well, how inhomogeneity in the size and positioning of the individual elements of the array affects its collective response. The results of this work contribute to improving the fundamental understanding of arrays of nanostructures, allowing for advancements in applications seeking to take advantage of their unique optical behavior.

\begin{figure*}
\begin{center}
\includegraphics[width=160mm,angle=0]{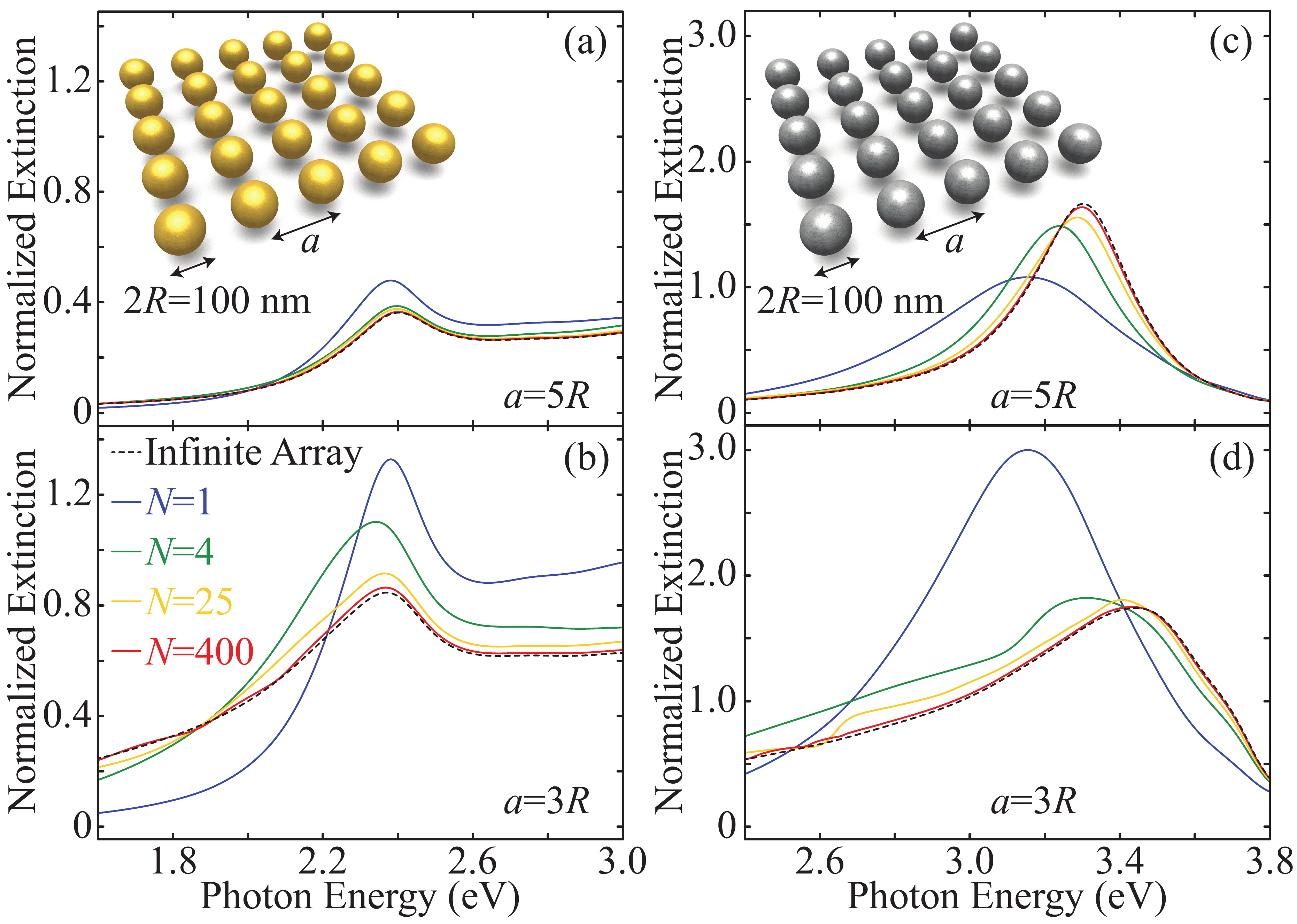}
\caption{Normalized extinction for arrays of metal nanospheres. Each array is composed of $N$ spheres of radius $R=50\,$nm arranged in a square lattice of period $a$, as shown in the insets of (a) and (c). (a,b) Extinction for arrays of gold nanospheres with $R=50\,$nm, and period $a=5R$ and $a=3R$, respectively. Blue curves correspond to the single particle case, while green, yellow, and red curves represent, respectively, arrays with $N=4$, $N=25$, and $N=400$. The black dashed curve shows the extinction for an infinite array. (c,d) Same as (a) and (b), but for silver nanospheres.} \label{fig1}
\end{center}
\end{figure*}

%\section{Results and Discussion}
The system under study consists of a self-standing square array with $N$ identical metallic nanospheres of radius $R$, separated by a center-to-center distance $a$, as shown by the insets of Figures~\ref{fig1}(a) and (c).  For all of the systems investigated in this work, we assume that the particles that constitute them are much smaller than the resonance wavelengths. This allows us to use a coupled dipole model to describe the optical response of the arrays, in which each constituent is characterized as an electric dipole with a scalar polarizability $\alpha$ \cite{ZKS03,paper090,C09,TD12,ama59}.  Our model, therefore, is valid for arrays of small metallic nanostructures, for which the contribution to their optical response arising from the magnetic dipole and higher order modes is negligible.
Upon illumination by an external field $\textbf{E}$, the induced dipole $\textbf{p}$ in each sphere satisfies
\begin{equation}\label{p}
\textbf{p}_i= \alpha \textbf{E}_i+\alpha\sum_{j\neq i} \textbf{G}_{ij} \textbf{p}_{j},
\end{equation}
where $\textbf{G}_{ij}$ is the interaction tensor that defines the interaction between dipoles $i$ and $j$
\begin{align}\label{g}
\textbf{G}_{ij}={}&\frac{e^{ikr_{ij}}}{r_{ij}^3}\left[\left(kr_{ij}\right)^2+ikr_{ij}-1\right]\mathbb{1}\nonumber \\ &-\frac{e^{ikr_{ij}}}{r_{ij}^3}\left[\left(kr_{ij}\right)^2+3ikr_{ij}-3\right]\frac{\textbf{r}_{ij}\otimes\textbf{r}_{ij}}{r_{ij}^2}.
\end{align}
Here, $\textbf{r}_{ij}=\textbf{r}_{i}-\textbf{r}_{j}$ is the vector connecting the positions of these dipoles, $\mathbb{1}$ is the $3\times3$ identity matrix, and $k=\omega/c=2\pi/\lambda$ is the wavenumber, with $\omega$ and $\lambda$ being, respectively, the frequency and the wavelength of light.
For a finite array, Eq.~(\ref{p}) can be solved as 
\begin{equation}\label{dipoles}
\textbf{p}_i= \sum^{N}_{j=1} [\alpha^{-1}\mathbb{1} - \textbf{G}]_{ij}^{-1} \textbf{E}_j,
\end{equation}
which involves the inversion of a $3N\times3N$ matrix. By doing so, we obtain the dipole induced at each particle, which, in turn, allows us to compute the extinction cross section of the array as
\begin{equation}\label{ext_fin}
\sigma_{\rm ext}=4\pi k \sum^{N}_{i=1} \textrm{Im}\{\textbf{p}_i\cdot \textbf{E}^{\ast}_i\}/ |\textbf{E}_i|^2.
\end{equation}
The polarizability of the spheres can be determined from the electric dipole Mie scattering coefficient $t_1^E$ as $\alpha=(3/2k^3)t_1^E$  \cite{paper112}, which is calculated analytically using the dielectric function of the material from which the nanoparticles are made. In this work, we use tabulated data taken from \cite{JC1972} to describe the dielectric functions of gold and silver. 

We can also use Eq.~(\ref{p}) to calculate the optical response of an infinite array. In this case, we take advantage of periodicity and use Bloch's theorem to write the external field and the induced dipoles as  $\textbf{E}_i=\textbf{E}(\textbf{k}_{\parallel})\exp(i\textbf{k}_{\parallel}\cdot\textbf{r}_i)$ and $\textbf{p}_i=\textbf{p}(\textbf{k}_{\parallel})\exp(i\textbf{k}_{\parallel}\cdot\textbf{r}_i)$, respectively, where $\textbf{k}_{\parallel}$ is the component of the wavevector of the incident field parallel to the array. By doing so, we get
\begin{equation*}\label{p_inf}
\textbf{p}(\textbf{k}_{\parallel})= \left[\alpha^{-1}\mathbb{1}- \mathcal{G}(\textbf{k}_{\parallel})\right]^{-1}\textbf{E}(\textbf{k}_{\parallel}),
\end{equation*}
where $\mathcal{G}(\textbf{k}_{\parallel})=\sum^{\infty}_{i\neq0}\textbf{G}_{i0}\exp(-i\textbf{k}_{\parallel}\cdot\textbf{r}_i)$. Once the induced dipole is calculated, the extinction cross section per unit cell is given by
\begin{equation}\label{ext_inf}
\sigma_{\rm ext}=4\pi k \textrm{Im}\{\textbf{p}(\textbf{k}_{\parallel})\cdot \textbf{E}^{*}(\textbf{k}_{\parallel})\}/|\textbf{E}(\textbf{k}_{\parallel})|^2.
\end{equation}

\begin{figure*}
\begin{center}
\includegraphics[width=160mm,angle=0]{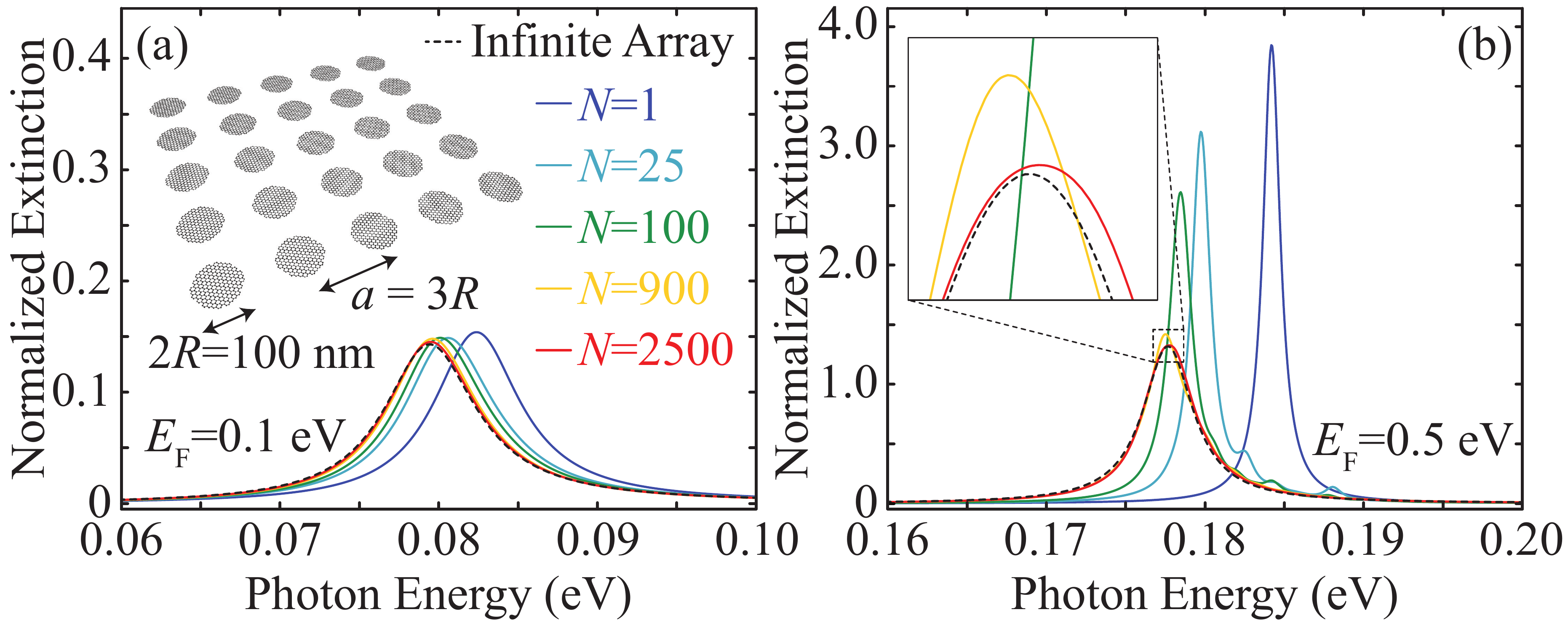}
\caption{Normalized extinction for arrays of graphene nanodisks. Each array consists of $N$ nanodisks of radius $R=50\,$nm arranged in a square lattice of period $a=3R$, as shown in the inset of (a). Blue curves display the results for the single particle case, while cyan, green, yellow, and red curves are used, respectively, to show those for arrays with $N=25$, $N=100$, $N=900$, and $N=2500$. The black dashed curve shows the extinction for an infinite array. For all of the cases, we consider two different doping levels: $E_{\rm F}=0.1\,$eV (a) and $E_{\rm F}=0.5\,$eV (b).   
The inset in (b) shows a zoom of the oscillations of the peak position for large $N$.
}\label{fig2}
\end{center}
\end{figure*} 

We use this model to calculate the extinction of a square array of gold nanospheres with radius $R=50\,$nm. The external field is always assumed to be normally incident on the array (\textit{i.e.}, $\textbf{k}_{\parallel}=0$) and polarized along one of its lattice vectors, which means that only the in-plane components of the dipoles are excited.
For the finite arrays, we normalize the extinction calculated from Eq.~(\ref{ext_fin}) by the area of the array, defined as $Na^2$, while, for the infinite array, we divide the output of Eq.~(\ref{ext_inf}) by the unit cell area $a^2$. The results of this calculation are shown in panel (a) of Figure~\ref{fig1} for an array with $a=5R$.
The blue curve corresponds to the single particle ($N=1$) case, while green, yellow, and red curves show the normalized extinction for arrays with increasing numbers of elements: $N=4$, $N=25$, and $N=400$, respectively. Examining these spectra, we observe that, as expected, when the number of particles increases, the extinction of the finite array approaches that of an infinite array (dashed black curve), becoming very similar to it for $4$ particles, and indistinguishable for $25$ particles. 
 The situation is different if the strength of the coupling between the nanoparticles is increased. This can be achieved by reducing the separation between them to $a=3R$, as done in panel (b). The corresponding results show a slower convergence, requiring up to $400$ particles to reach the infinite array limit. 

A similar behavior is obtained for arrays made of silver nanoparticles. In this case, the smaller losses and larger plasma frequency of silver give rise to stronger couplings, which result in a slightly slower convergence, as can be seen in panels (c) and (d), for $a=5R$ and $a=3R$, respectively. In both cases, we need to go up to $400$ particles to obtain a perfect agreement with the infinite array, although, for $a=5R$, the spectrum of the $N=25$ array is already very close to the infinite case. On the contrary, for $a=3R$, the spectrum of the $N=25$ array displays additional peaks at lower energies corresponding to higher order modes supported by the array due to its finite size.  

Interestingly, comparing the results for gold and silver arrays, we observe that the extinction peak for the latter clearly blueshifts as $N$ increases, which is different from the behavior displayed by the former. 
We attribute this difference to the role played by retardation in the response of these systems, as we explain later.

A stronger level of coupling can be achieved by substituting the metallic nanospheres for graphene nanodisks \cite{paper176}. When doped, these structures can support strong localized plasmon modes whose energy can be tuned by adjusting the doping level \cite{paper196}.  Due to these exceptional properties, arrays of graphene nanodisks have been proposed as a platform to develop tunable infrared plasmonic devices \cite{YLC12}, with functionalities such as total absorption \cite{paper182,paper230} and ultrasensitive biosensing \cite{paper256,ama53}, among many others \cite{paper196}.

In order to describe the response of these systems, we employ the coupled dipole model outlined above. In this case, however, the polarizability of the graphene nanodisks is calculated using 
\begin{equation*}\label{alpha_gra}
\alpha=\frac{8\omega_r^2R^3(-\eta)\zeta^2}{\omega_r^2-\omega^2-i\omega\gamma},
\end{equation*}
as derived within the Plasmon Wave Function (PWF) formalism \cite{paper235,paper303,ama53} using a Drude conductivity \cite{paper196}. Within that formalism, which is accurate for nanostructures with sizes much smaller than their resonance wavelength, the polarizability of an arbitrary nanodisk is characterized by the parameters $\eta=-0.0728$ and $\zeta=0.8508$, whose value are obtained by fitting the expression above to the rigorous solution of Maxwell's equations \cite{paper303}. These parameters, together with the radius $R$ of the disk and its Fermi energy $E_{\rm F}$, which quantifies the doping level of the nanodisk, define the frequency of the localized plasmon that it supports as
\begin{equation*}\label{wr}
\omega_r=\frac{e/\hbar}{\sqrt{-2\pi \eta}}\sqrt{\frac{E_{\rm F}}{R}}, 
\end{equation*}
while its linewidth is determined by $\gamma=ev_F^2/\mu E_{\rm F}$, with $v_F\approx c/300$ being the Fermi velocity of the electrons and $\mu$ their mobility, for which we assume a value of $10^4$ cm$^2/$(V s) \cite{paper176}.

Using these expressions, we compute the optical response of a square array of graphene nanodisks of radius $R=50\,$nm and period $a=3R$, as depicted in the inset of Figure~\ref{fig2}(a). This panel displays the extinction, normalized in the same way as in Figure~\ref{fig1}, for arrays with $E_{\rm F}=0.1\,$eV and different number of elements: $1$ (blue), $25$ (cyan), $100$  (green), $900$ (yellow), and $2500$ (red curves).
Examining these spectra, we observe that convergence to the infinite array limit, which is displayed by a black dashed curve, is reached for values of $N$ beyond $900$ nanodisks. This is the expected behavior, since the strong plasmons supported by graphene nanodisks lead to much stronger coupling between elements than those displayed by gold and silver nanostructures of the same size. 

The strength of the coupling between the nanodisks can be increased by raising its doping level, since, at resonance, $\alpha \propto \sqrt{E_{\rm F}}$. This can be seen in panel (b), where we plot the normalized extinction for an array identical to that of panel (a), but with $E_{\rm F}=0.5\,$eV. We observe that, in this case, the convergence to the infinite array limit requires a much larger number of elements ($N=2500$, red curve), consistent with the stronger level of coupling. Furthermore, smaller peaks, corresponding to higher order modes, are clearly visible on the righthand side of the main peak,  approaching to and eventually merging with it as $N$ increases.
Interestingly, looking at the inset, we observe that the peak for the $N=900$ array  (yellow curve) seems to have overshot that of the infinite array, shifting back when $N=2500$. This suggests that the process of convergence to the infinite array limit, as $N$ is increases, involves an oscillatory behavior. 

\begin{figure*}
\begin{center}
\includegraphics[width=160mm,angle=0]{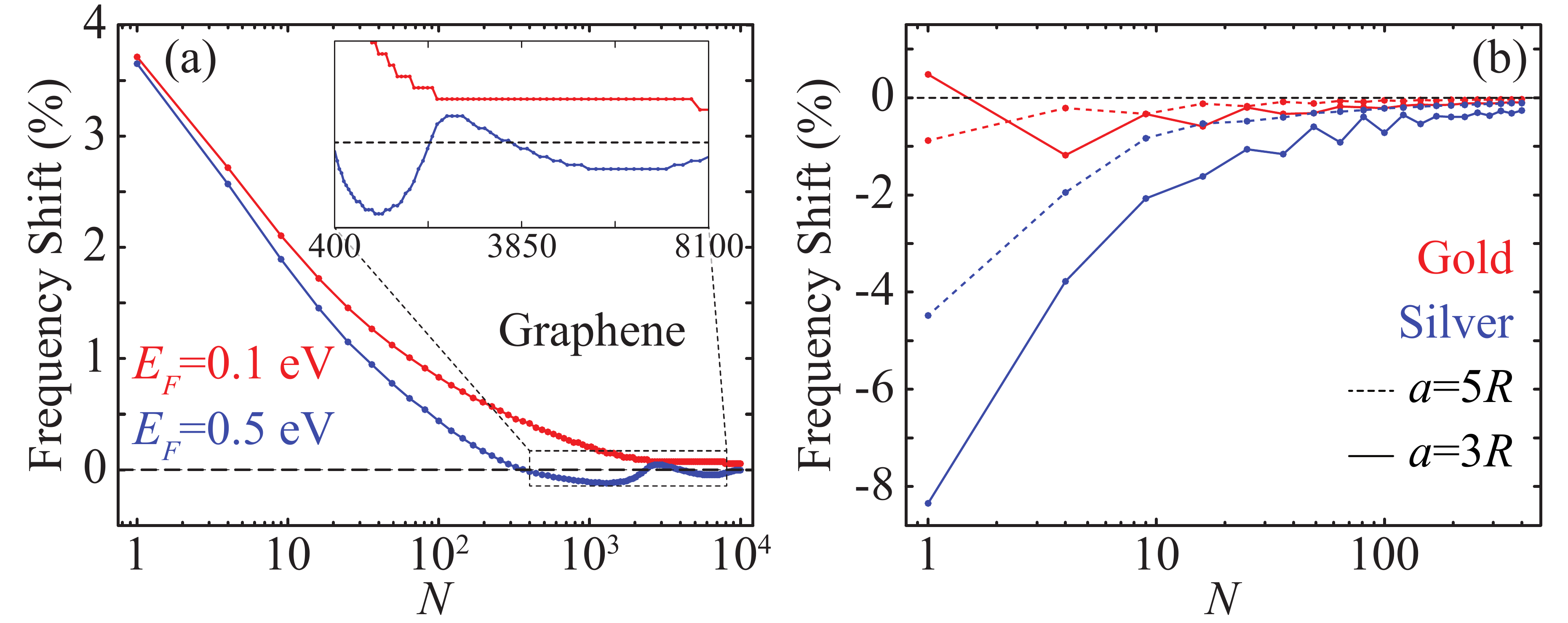}
\caption{ (a) Frequency shift of the extinction peak of finite arrays of graphene nanodisks measured with respect to that of an infinite one. Red and blue curves show the results for arrays with $E_{\rm F}=0.1\,$eV  and  $E_{\rm F}=0.5\,$eV, respectively. In all cases, $R=50\,$nm and $a=3R$. The inset shows a zoom of the $N=400-8100$ region. The discreteness in the results is a consequence of the finite energy resolution we use in the search of peak positions ($\approx 0.01\,$meV). (b) Frequency shifts for arrays of gold (red lines) and silver (blue lines) nanospheres of $R=50\,$nm, with periods $a=3R$ (solid lines) and $a=5R$ (dashed lines).} \label{fig3}
\end{center}\end{figure*}

We explore this behavior in Figure~\ref{fig3}(a), where we plot, as a function of $N$, the shift of the extinction peak with respect to that of the infinite array. Red and blue curves correspond, respectively, to $E_{\rm F}=0.1\,$eV and $E_{\rm F}=0.5\,$eV. These results clearly confirm the anticipated oscillatory behavior for $E_{\rm F}=0.5\,$eV, as can be seen in the inset, where we provide a zoom of the results for $N$ in the range $400-8100$. 
We attribute this behavior to the interplay between the main resonance and the higher order modes supported by the finite arrays as they converge to the spectrum of the infinite array when $N$ increases. 
In panel (b), we plot similar results for the arrays of metal nanospheres studied in Figure~\ref{fig1}. In this case, the red and blue colors are used for the gold and silver arrays, while solid and dashed curves correspond to $a=3R$ and $a=5R$, respectively. The weaker coupling in these systems, as compared with the graphene arrays, results in a faster convergence (\textit{cf.} the horizontal axes of panels (a) and (b)). However, they also show a similar oscillatory behavior, which went unnoticed in Figure~\ref{fig1} and that originates from the same phenomenon.

\begin{figure*}
\begin{center}
\includegraphics[width=160mm,angle=0]{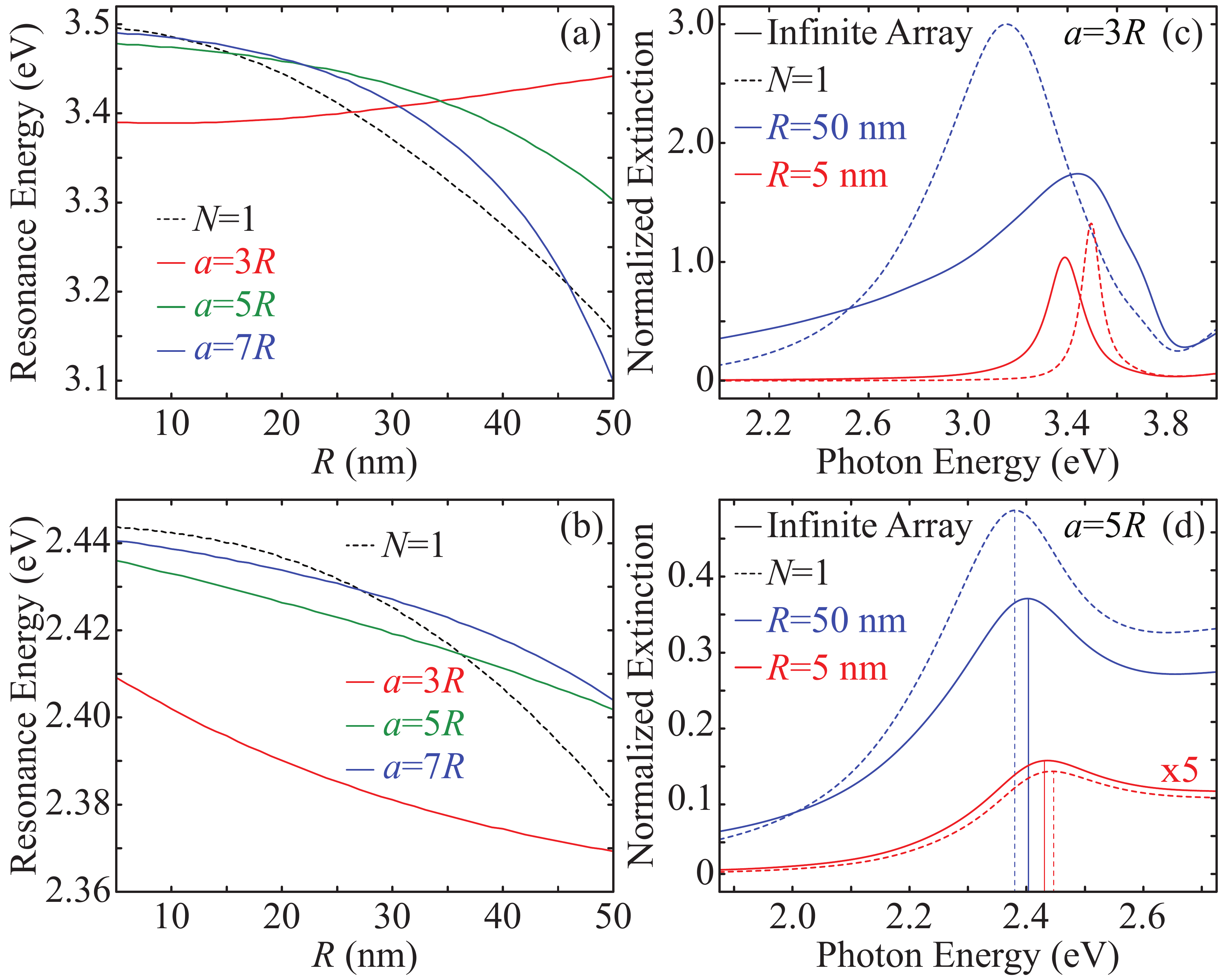}
\caption{(a) Extinction peak energy for infinite arrays of silver nanospheres as a function of the radius of the particles. The different solid curves corresponds to arrays with different periods: $a=3R$ (red curve), $a=5R$ (green curve), and $a=7R$ (blue curve), while the dashed curve shows the results for a single nanosphere. (b) Same as (a), but for gold nanospheres. (c) Normalized extinction for silver nanospheres with $R=5\,$nm (red curve) and $R=50\,$nm (blue curve). The solid curves correspond to the infinite array with $a=3R$, while the dashed curves represent the results for a single particle. (d) Same as (c), but for gold. In this case, the solid curves correspond to the infinite array with $a=5R$ and the red curves are multiplied by $5$ to improve visibility.} \label{fig4}
\end{center}
\end{figure*}

Figure~\ref{fig3} also shows that, when comparing the position of the extinction peak of the infinite array with that of a single particle, the graphene nanodisk arrays always show a redshift. 
In contrast to this, the behavior of the arrays composed of metallic nanospheres shows either a blue- or redshift, depending on the material and the period, as we noted in the discussion of Figure~\ref{fig1}.
This difference in behavior can be explained by analyzing the role played by retardation in the collective response of the arrays. 
The shift in the resonance of the array as $N$ increases is determined by the real part of the dipole-dipole interaction tensor (see Eq.~(\ref{dipoles})), and, in particular, the value it takes between nearest neighbors, for which the coupling is the strongest. A positive interaction results in a redshift of the collective resonance with respect to that of a single particle, while a negative one produces a blueshift.
 Examining Eq.~(\ref{g}), we infer that, when retardation is negligible (\textit{i.e.}, $ka\ll1$), this interaction reduces to $2a^{-3}$ for nearest neighbors, and therefore is positive. On the contrary, when retardation is significant (\textit{i.e.}, $ka \gtrsim 1$), the real part of the interaction becomes approximately $k^2\cos(ka)/a$, and, consequently, its sign depends on the particular value of $ka$. 
 
In order to confirm this explanation, we calculate the extinction peak energy for infinite arrays of silver and gold nanospheres with different radii and periods. The results of this calculation are shown in Figures~\ref{fig4}(a) and (b) using solid curves of different colors, as indicated by the legend. Examining panel (a) and  comparing these results with those for single nanospheres of the same size (black dashed curve), we observe that, in the case of silver nanoparticles, for $R=5\,$nm and $a=3R$, the resonance energy of the infinite array is clearly redshifted with respect to that of the single particle, as expected from the small value that  $ka$ takes for that case. However, if $R$ or $a$ is increased, the resonance energy of the infinite array approaches to that of the single particle and eventually crosses it, thus resulting in a blueshift. If $ka$ is further increased, the shift is reduced and it is possible to observe a second crossing, which is expected to happen when $ka\approx 3\pi/2$, for which $\cos(ka)$ changes its sign. The same trend is observed for gold nanoparticles, as seen from the results plotted in panel (b). In this case, however, the smaller plasma frequency of gold as compared with silver makes the crossings appear for larger values of $R$ and $a$. 

Figures~\ref{fig4}(c) and (d) show the normalized extinction spectra for two particular examples taken from panels (a) and (b), corresponding to $R=5\,$nm (red curves) and $R=50\,$nm (blue curves), with $a=3R$ in the case of silver and $a=5R$ for gold. The results for the infinite arrays are shown with solid curves, while those of the single particle are displayed using dashed curves. As anticipated, for $R=5\,$nm, the peak of the infinite array is redshifted with respect to that of the single particle, while, for $R=50\,$nm, the situation is reversed. It is worth noting that these predictions are in agreement with previous experimental observations \cite{SHV08}.

\begin{figure*}
\begin{center}
\includegraphics[width=120mm,angle=0]{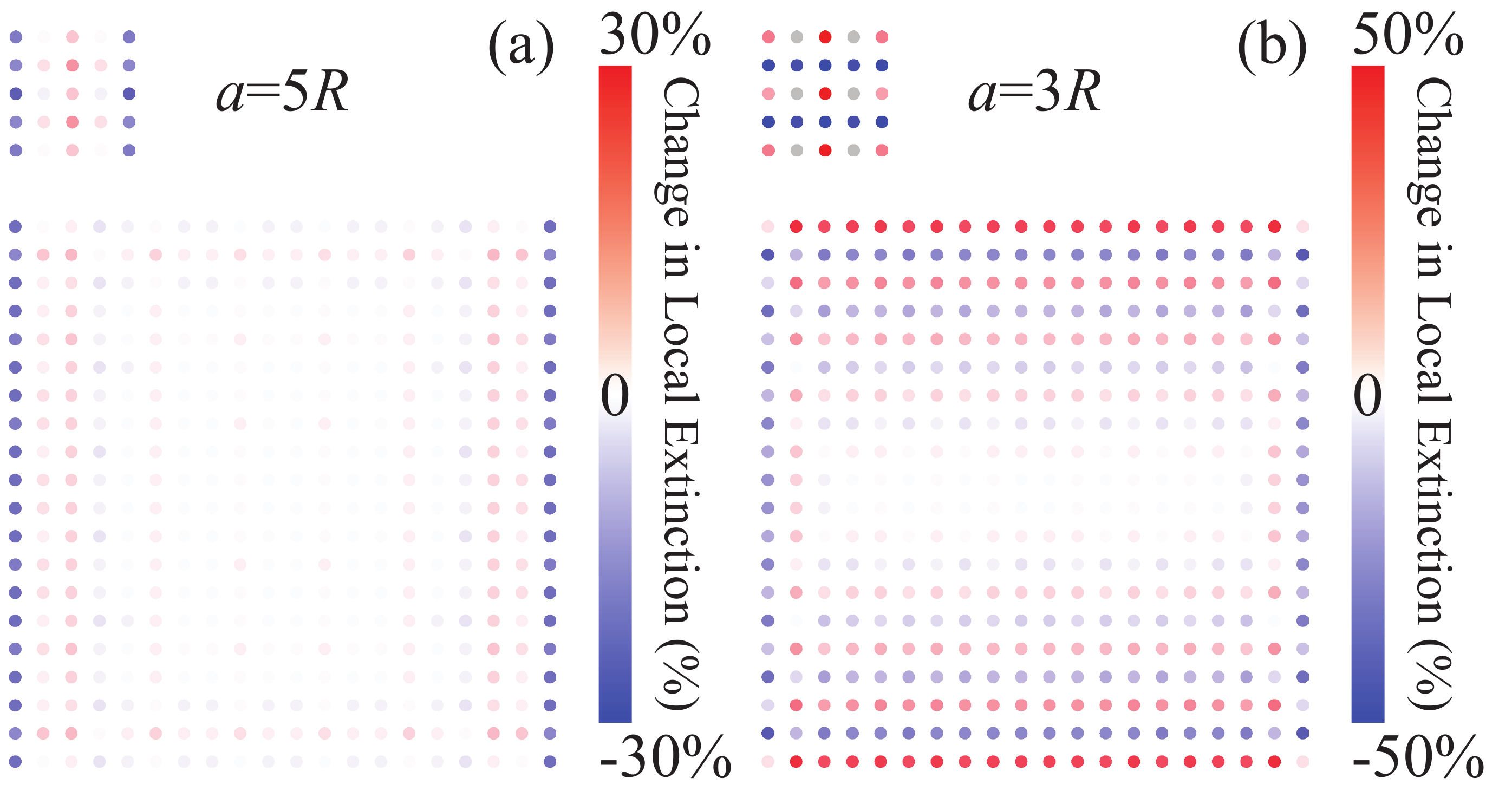}
\caption{Change in the local extinction of finite arrays of silver nanospheres with respect to the infinite array, calculated at the resonance frequency. We consider arrays composed of $N=25$ (upper plot) and $N=400$ (bottom plot) nanospheres of $R=50\,$nm, with a period of either $a=5R$ (a) or $a=3R$ (b). In all of the cases, we assume the illuminating field to be polarized along the vertical axis.} \label{fig5}
\end{center}
\end{figure*}

So far, we have gauged the convergence of the optical response of the finite arrays to the infinite array limit by analyzing their extinction, which is a quantity associated with the far-field response of the system. However, it is important to analyze as well how the near-field behavior of the array converges as $N$ increases. In particular, for an infinite array excited at normal incidence, the dipole induced at each particle has to be identical, due to the periodicity of the system. On the contrary, for finite arrays, the existence of edges breaks that symmetry. As a result of this, the dipoles induced at each particle may vary depending on its location within the array. In order to analyze this effect, we compute the local extinction produced by each dipole in the array at the resonance frequency, which is defined as $\sigma_{{\rm ext},i}=4\pi k {\rm Im}\{\textbf{p}_i\cdot \textbf{E}^{\ast}\}/|\textbf{E}_i|^2$.
Figure~\ref{fig5} shows the results of this calculation for different arrays of silver nanospheres, assuming an illuminating field polarized along the vertical axis. Specifically, we use colored circles to display the change in the local extinction with respect to the infinite array for each dipole in the array. We consider arrays of $R=50\,$nm nanospheres with $N=25$ (upper plot) and $N=400$ (lower plot) elements, with  period $a=5R$ (a) and $a=3R$ (b). As discussed in Figures~\ref{fig1}(c) and (d), all of these arrays have a total extinction that is very similar, if not identical, to that of the corresponding infinite array. However, examining the results of Figure~\ref{fig5}, we observe that, for certain particles, the local extinction shows variations as large as $\pm 50\%$ with respect to the value for the infinite array. As expected, these variations are more pronounced near the edges of the arrays, and for the systems with smaller period, for which the interaction between the elements is stronger.

\begin{figure*}
\begin{center}
\includegraphics[width=120mm,angle=0]{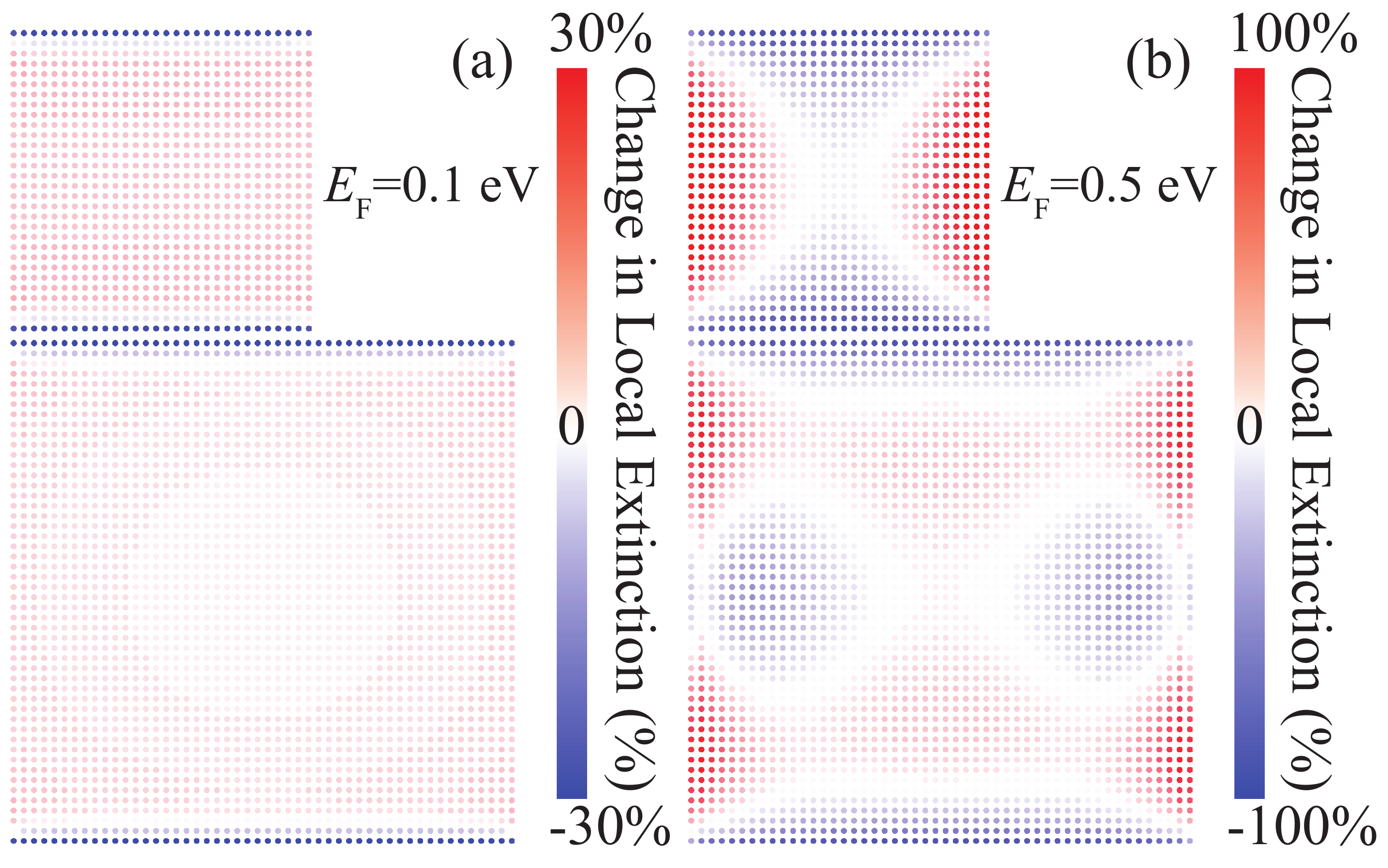}
\caption{Change in the local extinction of finite arrays of graphene nanodisks with respect to the infinite array, calculated at the resonance frequency. We consider arrays composed of $N=900$ (upper plot) and $N=2500$ (bottom plot) nanodisks of $R=50\,$nm, with period $a=3R$ and doping levels $E_{\rm F}=0.1\,$eV (a) and  $E_{\rm F}=0.5\,$eV (b).  In all of the cases, we assume the illuminating field to be polarized along the vertical axis.} \label{fig6}
\end{center}
\end{figure*}

Similar results are found for arrays of graphene nanodisks, as shown in Figure~\ref{fig6}. There, we plot the change in the local extinction with respect to the infinite array for systems with $R=50\,$nm and $a=3R$, and either $N=900$ (upper plot) or $N=2500$ (lower plot) nanodisks. As before, we assume the illuminating field to be polarized vertically. 
Panel (a) analyzes the results for $E_{\rm F}=0.1\,$eV, for which the change of the local extinction shows an approximately uniform pattern of positive values, except at the horizontal edges, where it turns negative, taking a value of almost $-30\%$. An increase in $E_{\rm F}$ results in more complicated patterns, as shown in panel (b), and larger variations up to $\pm 100\%$ of the infinite array value. These results demonstrate that, even if its extinction spectrum has already converged to the infinite array limit, the near-field response of a finite array can still show significant deviations from the infinite system behavior, especially  near the edges of the system.

\begin{figure*}
\begin{center}
\includegraphics[width=140mm,angle=0]{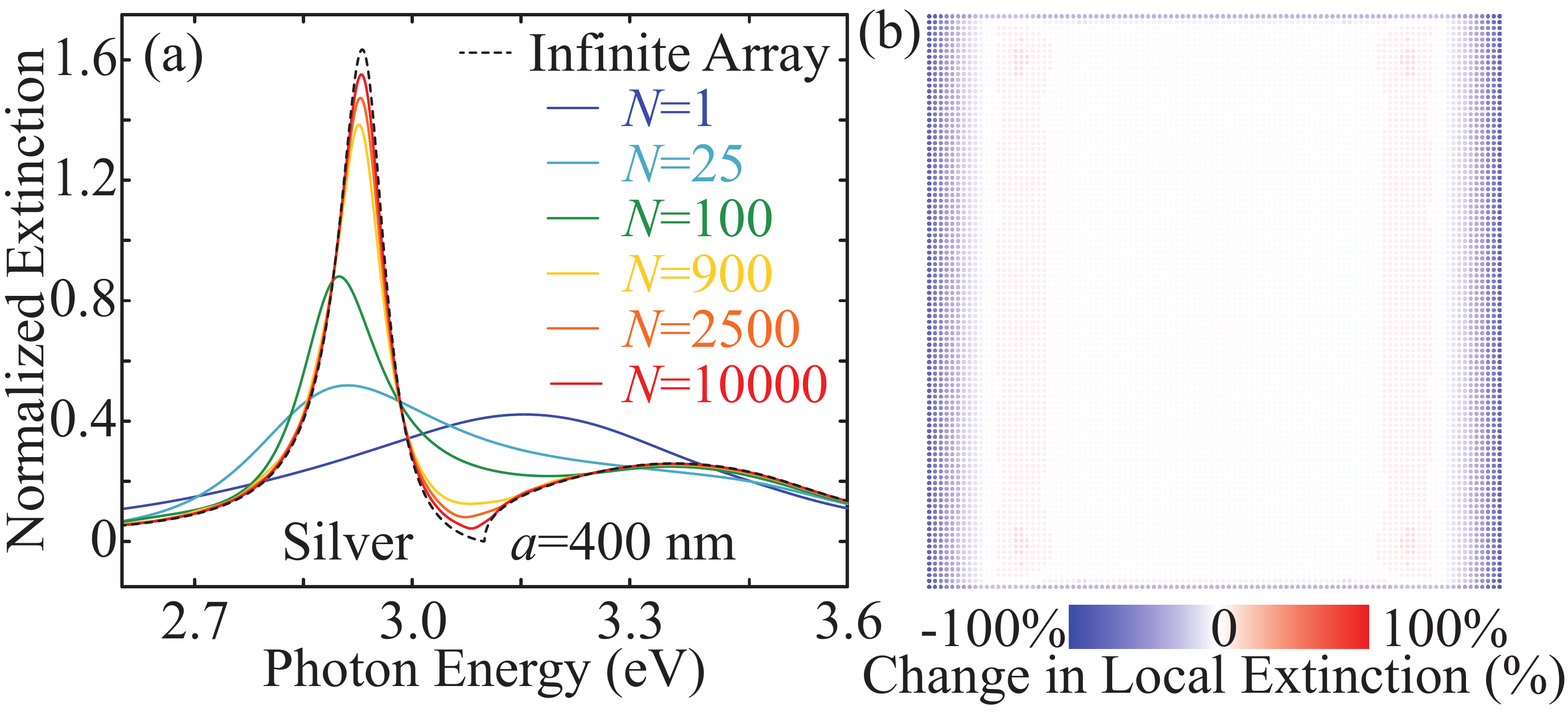}
\caption{(a) Normalized extinction for arrays of silver nanospheres having period $a=400\,$nm and radius of $R=50\,$nm. The color curves show the results for finite arrays with different number of elements, as indicated in the legend, while the black dashed curve displays the results for the corresponding infinite array. (b) Change in the local extinction with respect to the infinite array, calculated at the resonance frequency, for the array with $N=10000$.} \label{fig7}
\end{center}
\end{figure*}

In all of the analysis we have performed up to now, we have focused on arrays with periods smaller than their resonance wavelengths, for which the interaction between their elements is expected to be strong. Although, in principle, increasing the period is expected to lead to a smaller interaction, and, consequently, to a weaker collective response, there is an exception to this trend when the periodicity of the array is commensurate with the wavelength. In that case, the system can support the so-called lattice resonances, which arise from the coherent interaction of all of the elements of the system \cite{TD12,ama59}. 
In contrast to the resonances displayed by the systems we have analyzed above, which arise from the interaction between the plasmonic modes supported by  the constituents, the lattice resonances have a geometrical origin, and therefore are expected to be more sensitive to finite size effects \cite{MMT17}. In order to analyze this, we calculate the extinction produced by arrays of silver nanospheres with radius $50\,$nm, period $400\,$nm, and different number of elements. The corresponding normalized extinction spectra are shown in Figure~\ref{fig7}(a). Specifically, we consider arrays with $N$ ranging from $1$ to $10000$ (color curves), which we compare with the corresponding extinction for an infinite array (black dashed curves). 
The results confirm that convergence to the infinite array limit is significantly slower for these systems than for any of the cases investigated before. In particular, it is necessary to increase $N$ up to $10000$ to obtain a spectrum resembling that of the infinite array. However, even for that large number of elements, the characteristic sharp dip of the lattice resonance, associated with its Fano character \cite{paper090}, is not completely recovered.  Furthermore, as $N$ increases, the resonance peak becomes narrower \cite{FPP10,JR12}.
To complete our analysis, we plot, in panel (b), the change in the local extinction with respect to the infinite array for the system with $N=10000$ nanospheres,  calculated at resonance. Once again, we observe significant variations located near the edges of the array, although, in this case, there is a large central region for which the local extinction is almost identical to that of the infinite array.

All of the calculations discussed so far assume ideal arrays, in which all of the nanostructures have the same radius $R$ and are located at the exact positions defined by a square lattice of period $a$.
However, due to fabrication imperfections, any experimental realization of these systems will present defects, resulting in both the size of the particles and the separation between them having a certain finite distribution of values around the design values. 
In order to quantify how these defects impact the behavior of the system, we study the optical response of arrays of silver nanoparticles with inhomogeneities in the size and positioning of each element within the array. 
In particular, we build these arrays by adding $\delta R$ to the radius of each particle in the system, where $\delta R$ is a randomly generated number taken from the interval $[-\beta R,\beta R]$, with $\beta$ being a parameter that defines the level of disorder. Similarly, we shift the position of each particle by adding $(\delta x,\delta y)$ to its coordinates, where $\delta x$ and $\delta y$ are random numbers in the interval $[-\beta a,\beta a]$.

\begin{figure*}
\begin{center}
\includegraphics[width=160mm,angle=0]{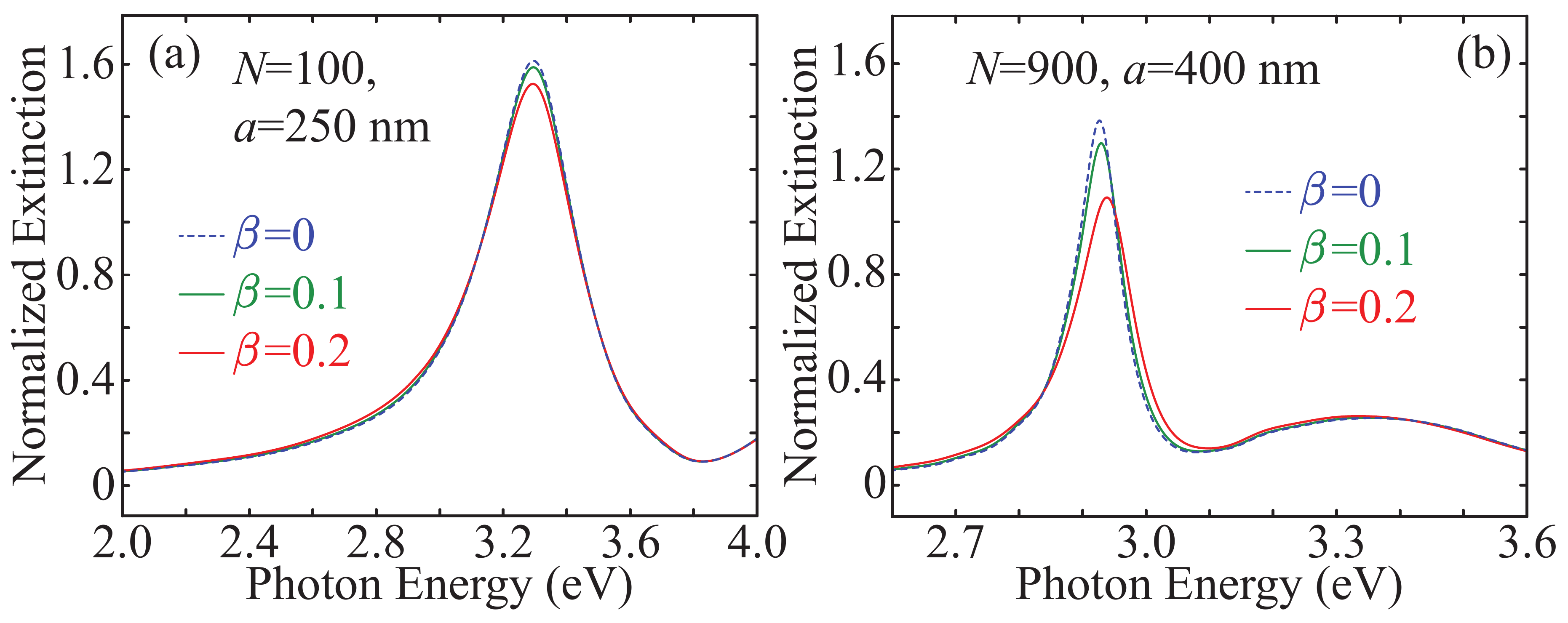}
\caption{Effect of inhomogeneity on finite arrays of silver nanospheres. (a) Extinction spectrum of an array with $a=250\,$nm composed of $N=100$ silver nanospheres with $R=50\,$nm. The dashed blue curve corresponds to an idealized array, for which the placement and size of each element is precise, while the green solid curve shows the extinction assuming $10\%$ inhomogeneity ($\beta=0.1$), and the red curve shows it for $\beta=0.2$ (\textit{i.e}, $20\%$ inhomogeneity). (b) Same as (a), but for $N=900$ nanospheres and a period $a=400\,$nm.} \label{fig8}
\end{center}
\end{figure*}

For each value of $\beta$, we perform $20$ different calculations of the extinction, each with their own randomly generated values of $\delta R$, $\delta x$, and $\delta y$ for each element in the array.
We plot the average of these runs in Figure~\ref{fig8} and compare it to the extinction for the array having a perfectly precise placement and size (blue dashed curves). In panel (a), this is done for an array of $N=100$ nanospheres of radius $R=50\,$nm with a period $a=250\,$nm, which is smaller than the resonance wavelength. We observe that, for a value of $\beta=0.1$ (green curve), which corresponds to a $10\%$ inhomogeneity in size and position, the extinction remains virtually unchanged, with only a slight broadening of the peak and a small decrease in the maximum extinction value. This is also the case  when the deviations in the size and position of the particles are allowed to reach $20\%$ (\textit{i.e.}, $\beta=0.2$) of their nominal values, as shown by the red curve. 

A different situation is found when the period of the array is increased to be similar to the resonance wavelength. In this case, the effect of inhomogeneity on the extinction is more pronounced. This can be seen in panel (b), where we study an array of $N=900$ silver nanospheres with $a=400\,$nm and $R=50\,$nm. The extinction of this array is slightly changed when $\beta=0.1$, with its peak reaching approximately $94\%$ of that of the ideal array. However, when $\beta$ is further increased to $0.2$, the extinction is changed significantly, with the extinction peak dropping to $79\%$ of the maximum for the ideal system. This behavior is not surprising, since the collective nature of lattice resonances makes them more sensitive to disorder \cite{PFF09,AB09,SK15}.

%\section{Conclusions}

In conclusion, we have analyzed of the evolution of the optical response of finite arrays of nanostructures as their number of elements is increased, and approaches the perfectly periodic, infinite array limit. 
Using a coupled dipole model, we have investigated arrays of gold and silver nanospheres, as well as graphene nanodisks, with periods smaller than their resonance wavelength. We have found that the number of elements required for convergence to the infinite array limit depends heavily on the strength of the coupling between each element. Furthermore, 
the evolution of the optical response of these systems, as it converges to the infinite array limit, is strongly dependent on the role played by retardation. In particular, when retardation is not significant, there is always a redshift with increasing number of elements, whereas, in the opposite limit, we predict either a blueshift or a redshift, depending on the particular values of the resonance wavelength and the array period. 
We have also found that, for relevant structures, even when their far-field responses may have converged to the infinite array limit, the near-field properties of the system can display significant inhomogeneities, which are especially significant at the edges of the array. 

We have also investigated the finite-size effects on the optical response of arrays with periods similar to their resonance wavelength, which can support lattice resonances arising from the coherent coupling of all of their elements enabled by their periodicity. We have found that, for these systems, due to the geometrical origin of these resonances, the convergence to the infinite array limit requires a significantly larger number of elements. 
Finally, we have performed a detailed analysis of the effect that the disorder in the position of the nanoparticles and the inhomogeneity in their sizes have on the optical response of different arrays of nanostructures. We have shown that, while arrays with periods smaller than the resonance wavelength show a significant robustness against disorder, systems supporting lattice resonances are more sensitive to it.

The results presented here provide a comprehensive analysis of the impact that finite-size effects have on the optical response of periodic arrays of nanostructures, thus contributing to the fundamental understanding of these systems and laying the foundations for future applications exploiting their unique optical properties.

\begin{acknowledgments}
This work has been sponsored by the U.S. National Science Foundation (Grant ECCS-1710697). The authors acknowledge the UNM Center for Advanced Research Computing for the computational resources used in this work. L.Z. acknowledges support from the Rayburn Reaching Up Fund and the New Mexico Space Grant Consortium. 
\end{acknowledgments}

%\bibliographystyle{apsrev}
%\bibliography{../../../refs}

\begin{thebibliography}{63}
\expandafter\ifx\csname natexlab\endcsname\relax\def\natexlab#1{#1}\fi
\expandafter\ifx\csname bibnamefont\endcsname\relax
  \def\bibnamefont#1{#1}\fi
\expandafter\ifx\csname bibfnamefont\endcsname\relax
  \def\bibfnamefont#1{#1}\fi
\expandafter\ifx\csname citenamefont\endcsname\relax
  \def\citenamefont#1{#1}\fi
\expandafter\ifx\csname url\endcsname\relax
  \def\url#1{\texttt{#1}}\fi
\expandafter\ifx\csname urlprefix\endcsname\relax\def\urlprefix{URL }\fi
\providecommand{\bibinfo}[2]{#2}
\providecommand{\eprint}[2][]{\url{#2}}

\bibitem[{\citenamefont{Maier}(2007)}]{M07}
\bibinfo{author}{\bibfnamefont{S.~A.} \bibnamefont{Maier}},
  \emph{\bibinfo{title}{Plasmonics: Fundamentals and Applications}}
  (\bibinfo{publisher}{Springer}, \bibinfo{address}{New York},
  \bibinfo{year}{2007}).

\bibitem[{\citenamefont{Xu et~al.}(1999)\citenamefont{Xu, Bjerneld, {K\"all},
  and {B\"orjesson}}}]{XBK99}
\bibinfo{author}{\bibfnamefont{H.}~\bibnamefont{Xu}},
  \bibinfo{author}{\bibfnamefont{E.~J.} \bibnamefont{Bjerneld}},
  \bibinfo{author}{\bibfnamefont{M.}~\bibnamefont{{K\"all}}}, \bibnamefont{and}
  \bibinfo{author}{\bibfnamefont{L.}~\bibnamefont{{B\"orjesson}}},
  \bibinfo{journal}{Phys.\ Rev.\ Lett.} \textbf{\bibinfo{volume}{83}},
  \bibinfo{pages}{4357} (\bibinfo{year}{1999}).

\bibitem[{\citenamefont{Anker et~al.}(2008)\citenamefont{Anker, Hall, Lyandres,
  Shah, Zhao, and {Van Duyne}}}]{AHL08}
\bibinfo{author}{\bibfnamefont{J.~N.} \bibnamefont{Anker}},
  \bibinfo{author}{\bibfnamefont{W.~P.} \bibnamefont{Hall}},
  \bibinfo{author}{\bibfnamefont{O.}~\bibnamefont{Lyandres}},
  \bibinfo{author}{\bibfnamefont{N.~C.} \bibnamefont{Shah}},
  \bibinfo{author}{\bibfnamefont{J.}~\bibnamefont{Zhao}}, \bibnamefont{and}
  \bibinfo{author}{\bibfnamefont{R.~P.} \bibnamefont{{Van Duyne}}},
  \bibinfo{journal}{Nat.\ Mater.} \textbf{\bibinfo{volume}{7}},
  \bibinfo{pages}{442} (\bibinfo{year}{2008}).

\bibitem[{\citenamefont{Catchpole and Polman}(2008)}]{CP08}
\bibinfo{author}{\bibfnamefont{K.~R.} \bibnamefont{Catchpole}}
  \bibnamefont{and} \bibinfo{author}{\bibfnamefont{A.}~\bibnamefont{Polman}},
  \bibinfo{journal}{Opt.\ Express} \textbf{\bibinfo{volume}{16}},
  \bibinfo{pages}{21793} (\bibinfo{year}{2008}).

\bibitem[{\citenamefont{Atwater and Polman}(2010)}]{AP10}
\bibinfo{author}{\bibfnamefont{H.~A.} \bibnamefont{Atwater}} \bibnamefont{and}
  \bibinfo{author}{\bibfnamefont{A.}~\bibnamefont{Polman}},
  \bibinfo{journal}{Nat.\ Mater.} \textbf{\bibinfo{volume}{9}},
  \bibinfo{pages}{205} (\bibinfo{year}{2010}).

\bibitem[{\citenamefont{Baffou and Quidant}(2014)}]{BQ14}
\bibinfo{author}{\bibfnamefont{G.}~\bibnamefont{Baffou}} \bibnamefont{and}
  \bibinfo{author}{\bibfnamefont{R.}~\bibnamefont{Quidant}},
  \bibinfo{journal}{Chem. Soc. Rev.} pp. \bibinfo{pages}{3898--3907}
  (\bibinfo{year}{2014}).

\bibitem[{\citenamefont{Brongersma et~al.}(2015)\citenamefont{Brongersma,
  Halas, and Nordlander}}]{BHN15}
\bibinfo{author}{\bibfnamefont{M.~L.} \bibnamefont{Brongersma}},
  \bibinfo{author}{\bibfnamefont{N.~J.} \bibnamefont{Halas}}, \bibnamefont{and}
  \bibinfo{author}{\bibfnamefont{P.}~\bibnamefont{Nordlander}},
  \bibinfo{journal}{Nat.\ Nanotechnol.} \textbf{\bibinfo{volume}{10}},
  \bibinfo{pages}{25} (\bibinfo{year}{2015}).

\bibitem[{\citenamefont{Adamo et~al.}(2012)\citenamefont{Adamo, Ou, So,
  Jenkins, De~Angelis, MacDonald, Di~Fabrizio, Ruostekoski, and
  Zheludev}}]{AOS12}
\bibinfo{author}{\bibfnamefont{G.}~\bibnamefont{Adamo}},
  \bibinfo{author}{\bibfnamefont{J.~Y.} \bibnamefont{Ou}},
  \bibinfo{author}{\bibfnamefont{J.~K.} \bibnamefont{So}},
  \bibinfo{author}{\bibfnamefont{S.~D.} \bibnamefont{Jenkins}},
  \bibinfo{author}{\bibfnamefont{F.}~\bibnamefont{De~Angelis}},
  \bibinfo{author}{\bibfnamefont{K.~F.} \bibnamefont{MacDonald}},
  \bibinfo{author}{\bibfnamefont{E.}~\bibnamefont{Di~Fabrizio}},
  \bibinfo{author}{\bibfnamefont{J.}~\bibnamefont{Ruostekoski}},
  \bibnamefont{and} \bibinfo{author}{\bibfnamefont{N.~I.}
  \bibnamefont{Zheludev}}, \bibinfo{journal}{Phys.\ Rev.\ Lett.}
  \textbf{\bibinfo{volume}{109}}, \bibinfo{pages}{217401}
  (\bibinfo{year}{2012}).

\bibitem[{\citenamefont{Lozano et~al.}(2013)\citenamefont{Lozano, Louwers,
  Rodriguez, Murai, Jansen, Verschuuren, and {Gomez Rivas}}}]{LLR13}
\bibinfo{author}{\bibfnamefont{G.}~\bibnamefont{Lozano}},
  \bibinfo{author}{\bibfnamefont{D.~J.} \bibnamefont{Louwers}},
  \bibinfo{author}{\bibfnamefont{S.~R.~K.} \bibnamefont{Rodriguez}},
  \bibinfo{author}{\bibfnamefont{S.}~\bibnamefont{Murai}},
  \bibinfo{author}{\bibfnamefont{O.~T.~A.} \bibnamefont{Jansen}},
  \bibinfo{author}{\bibfnamefont{M.~A.} \bibnamefont{Verschuuren}},
  \bibnamefont{and} \bibinfo{author}{\bibfnamefont{J.}~\bibnamefont{{Gomez
  Rivas}}}, \bibinfo{journal}{Light\ Sci.\ Appl.} \textbf{\bibinfo{volume}{2}},
  \bibinfo{pages}{e241} (\bibinfo{year}{2013}).

\bibitem[{\citenamefont{Zhou et~al.}(2013)\citenamefont{Zhou, Dridi, Suh, Kim,
  Co, Wasielewski, Schatz, and Odom}}]{ZDS13}
\bibinfo{author}{\bibfnamefont{W.}~\bibnamefont{Zhou}},
  \bibinfo{author}{\bibfnamefont{M.}~\bibnamefont{Dridi}},
  \bibinfo{author}{\bibfnamefont{J.~Y.} \bibnamefont{Suh}},
  \bibinfo{author}{\bibfnamefont{C.~H.} \bibnamefont{Kim}},
  \bibinfo{author}{\bibfnamefont{D.~T.} \bibnamefont{Co}},
  \bibinfo{author}{\bibfnamefont{M.~R.} \bibnamefont{Wasielewski}},
  \bibinfo{author}{\bibfnamefont{G.~C.} \bibnamefont{Schatz}},
  \bibnamefont{and} \bibinfo{author}{\bibfnamefont{T.~W.} \bibnamefont{Odom}},
  \bibinfo{journal}{Nat.\ Nanotechnol.} \textbf{\bibinfo{volume}{8}},
  \bibinfo{pages}{506} (\bibinfo{year}{2013}).

\bibitem[{\citenamefont{Yang and Odom}(2015)}]{YO15}
\bibinfo{author}{\bibfnamefont{A.}~\bibnamefont{Yang}} \bibnamefont{and}
  \bibinfo{author}{\bibfnamefont{T.~W.} \bibnamefont{Odom}},
  \bibinfo{journal}{IEEE Photonics Journal} \textbf{\bibinfo{volume}{7}},
  \bibinfo{pages}{1} (\bibinfo{year}{2015}).

\bibitem[{\citenamefont{Hakala et~al.}(2017)\citenamefont{Hakala, Rekola,
  V\"akev\"ainen, Martikainen, Ne\v{c}ada, Moilanen, and T\"orm\"a}}]{HRV17}
\bibinfo{author}{\bibfnamefont{T.~K.} \bibnamefont{Hakala}},
  \bibinfo{author}{\bibfnamefont{H.~T.} \bibnamefont{Rekola}},
  \bibinfo{author}{\bibfnamefont{A.~I.} \bibnamefont{V\"akev\"ainen}},
  \bibinfo{author}{\bibfnamefont{J.-P.} \bibnamefont{Martikainen}},
  \bibinfo{author}{\bibfnamefont{M.}~\bibnamefont{Ne\v{c}ada}},
  \bibinfo{author}{\bibfnamefont{A.~J.} \bibnamefont{Moilanen}},
  \bibnamefont{and}
  \bibinfo{author}{\bibfnamefont{P.}~\bibnamefont{T\"orm\"a}},
  \bibinfo{journal}{Nat.\ Commun.} \textbf{\bibinfo{volume}{8}},
  \bibinfo{pages}{13687} (\bibinfo{year}{2017}).

\bibitem[{\citenamefont{Alitalo et~al.}(2006)\citenamefont{Alitalo, Simovski,
  Viitanen, and Tretyakov}}]{ASV06}
\bibinfo{author}{\bibfnamefont{P.}~\bibnamefont{Alitalo}},
  \bibinfo{author}{\bibfnamefont{C.}~\bibnamefont{Simovski}},
  \bibinfo{author}{\bibfnamefont{A.}~\bibnamefont{Viitanen}}, \bibnamefont{and}
  \bibinfo{author}{\bibfnamefont{S.}~\bibnamefont{Tretyakov}},
  \bibinfo{journal}{Phys.\ Rev.\ B} \textbf{\bibinfo{volume}{74}},
  \bibinfo{pages}{235425} (\bibinfo{year}{2006}).

\bibitem[{\citenamefont{Willets et~al.}(2017)\citenamefont{Willets, Wilson,
  Sundaresan, and Joshi}}]{WWS17}
\bibinfo{author}{\bibfnamefont{K.~A.} \bibnamefont{Willets}},
  \bibinfo{author}{\bibfnamefont{A.~J.} \bibnamefont{Wilson}},
  \bibinfo{author}{\bibfnamefont{V.}~\bibnamefont{Sundaresan}},
  \bibnamefont{and} \bibinfo{author}{\bibfnamefont{P.~B.} \bibnamefont{Joshi}},
  \bibinfo{journal}{Chem.\ Rev.} \textbf{\bibinfo{volume}{117}},
  \bibinfo{pages}{7538} (\bibinfo{year}{2017}).

\bibitem[{\citenamefont{Noskov et~al.}(2012)\citenamefont{Noskov, Belov, and
  Kivshar}}]{NBK12}
\bibinfo{author}{\bibfnamefont{R.~E.} \bibnamefont{Noskov}},
  \bibinfo{author}{\bibfnamefont{P.~A.} \bibnamefont{Belov}}, \bibnamefont{and}
  \bibinfo{author}{\bibfnamefont{Y.~S.} \bibnamefont{Kivshar}},
  \bibinfo{journal}{Opt.\ Express} \textbf{\bibinfo{volume}{20}},
  \bibinfo{pages}{2733} (\bibinfo{year}{2012}).

\bibitem[{\citenamefont{Fan et~al.}(2006)\citenamefont{Fan, Zhang, Panoiu,
  Abdenour, Krishna, Osgood, Malloy, and Brueck}}]{FZP06}
\bibinfo{author}{\bibfnamefont{W.}~\bibnamefont{Fan}},
  \bibinfo{author}{\bibfnamefont{S.}~\bibnamefont{Zhang}},
  \bibinfo{author}{\bibfnamefont{N.-C.} \bibnamefont{Panoiu}},
  \bibinfo{author}{\bibfnamefont{A.}~\bibnamefont{Abdenour}},
  \bibinfo{author}{\bibfnamefont{S.}~\bibnamefont{Krishna}},
  \bibinfo{author}{\bibfnamefont{R.~M.} \bibnamefont{Osgood}},
  \bibinfo{author}{\bibfnamefont{K.~J.} \bibnamefont{Malloy}},
  \bibnamefont{and} \bibinfo{author}{\bibfnamefont{S.~R.~J.}
  \bibnamefont{Brueck}}, \bibinfo{journal}{Nano\ Lett.}
  \textbf{\bibinfo{volume}{6}}, \bibinfo{pages}{1027} (\bibinfo{year}{2006}).

\bibitem[{\citenamefont{Metzger et~al.}(2015)\citenamefont{Metzger, Gui, Fuchs,
  Floess, Hentschel, and Giessen}}]{MGF15}
\bibinfo{author}{\bibfnamefont{B.}~\bibnamefont{Metzger}},
  \bibinfo{author}{\bibfnamefont{L.}~\bibnamefont{Gui}},
  \bibinfo{author}{\bibfnamefont{J.}~\bibnamefont{Fuchs}},
  \bibinfo{author}{\bibfnamefont{D.}~\bibnamefont{Floess}},
  \bibinfo{author}{\bibfnamefont{M.}~\bibnamefont{Hentschel}},
  \bibnamefont{and} \bibinfo{author}{\bibfnamefont{H.}~\bibnamefont{Giessen}},
  \bibinfo{journal}{Nano\ Lett.} \textbf{\bibinfo{volume}{15}},
  \bibinfo{pages}{3917} (\bibinfo{year}{2015}).

\bibitem[{\citenamefont{Yu and Capasso}(2014)}]{YC14}
\bibinfo{author}{\bibfnamefont{N.}~\bibnamefont{Yu}} \bibnamefont{and}
  \bibinfo{author}{\bibfnamefont{F.}~\bibnamefont{Capasso}},
  \bibinfo{journal}{Nat.\ Mater.} \textbf{\bibinfo{volume}{13}},
  \bibinfo{pages}{139} (\bibinfo{year}{2014}).

\bibitem[{\citenamefont{Yu et~al.}(2011)\citenamefont{Yu, Genevet, Kats, Aieta,
  Tetienne, Capasso, and Gaburro}}]{YGK11}
\bibinfo{author}{\bibfnamefont{N.}~\bibnamefont{Yu}},
  \bibinfo{author}{\bibfnamefont{P.}~\bibnamefont{Genevet}},
  \bibinfo{author}{\bibfnamefont{M.~A.} \bibnamefont{Kats}},
  \bibinfo{author}{\bibfnamefont{F.}~\bibnamefont{Aieta}},
  \bibinfo{author}{\bibfnamefont{J.-P.} \bibnamefont{Tetienne}},
  \bibinfo{author}{\bibfnamefont{F.}~\bibnamefont{Capasso}}, \bibnamefont{and}
  \bibinfo{author}{\bibfnamefont{Z.}~\bibnamefont{Gaburro}},
  \bibinfo{journal}{Science} \textbf{\bibinfo{volume}{334}},
  \bibinfo{pages}{333} (\bibinfo{year}{2011}).

\bibitem[{\citenamefont{Wang et~al.}(2018)\citenamefont{Wang, Ramezani,
  V\"akev\"ainen, T\"orm\"a, {G\'omez Rivas}, and Odom}}]{WRV18}
\bibinfo{author}{\bibfnamefont{W.}~\bibnamefont{Wang}},
  \bibinfo{author}{\bibfnamefont{M.}~\bibnamefont{Ramezani}},
  \bibinfo{author}{\bibfnamefont{A.~I.} \bibnamefont{V\"akev\"ainen}},
  \bibinfo{author}{\bibfnamefont{P.}~\bibnamefont{T\"orm\"a}},
  \bibinfo{author}{\bibfnamefont{J.}~\bibnamefont{{G\'omez Rivas}}},
  \bibnamefont{and} \bibinfo{author}{\bibfnamefont{T.~W.} \bibnamefont{Odom}},
  \bibinfo{journal}{Mater.\ Today} \textbf{\bibinfo{volume}{21}},
  \bibinfo{pages}{303 } (\bibinfo{year}{2018}).

\bibitem[{\citenamefont{{Garc\'{\i}a de Abajo}}(2007)}]{paper090}
\bibinfo{author}{\bibfnamefont{F.~J.} \bibnamefont{{Garc\'{\i}a de Abajo}}},
  \bibinfo{journal}{Rev.\ Mod.\ Phys.} \textbf{\bibinfo{volume}{79}},
  \bibinfo{pages}{1267} (\bibinfo{year}{2007}).

\bibitem[{\citenamefont{Jenkins et~al.}(2017)\citenamefont{Jenkins,
  Ruostekoski, Papasimakis, Savo, and Zheludev}}]{JRP17}
\bibinfo{author}{\bibfnamefont{S.~D.} \bibnamefont{Jenkins}},
  \bibinfo{author}{\bibfnamefont{J.}~\bibnamefont{Ruostekoski}},
  \bibinfo{author}{\bibfnamefont{N.}~\bibnamefont{Papasimakis}},
  \bibinfo{author}{\bibfnamefont{S.}~\bibnamefont{Savo}}, \bibnamefont{and}
  \bibinfo{author}{\bibfnamefont{N.~I.} \bibnamefont{Zheludev}},
  \bibinfo{journal}{Phys.\ Rev.\ Lett.} \textbf{\bibinfo{volume}{119}},
  \bibinfo{pages}{053901} (\bibinfo{year}{2017}).

\bibitem[{\citenamefont{Hasan et~al.}(2018)\citenamefont{Hasan, Mosk, Vos, and
  Lagendijk}}]{HMV18}
\bibinfo{author}{\bibfnamefont{S.~B.} \bibnamefont{Hasan}},
  \bibinfo{author}{\bibfnamefont{A.~P.} \bibnamefont{Mosk}},
  \bibinfo{author}{\bibfnamefont{W.~L.} \bibnamefont{Vos}}, \bibnamefont{and}
  \bibinfo{author}{\bibfnamefont{A.}~\bibnamefont{Lagendijk}},
  \bibinfo{journal}{Phys.\ Rev.\ Lett.} \textbf{\bibinfo{volume}{120}},
  \bibinfo{pages}{237402} (\bibinfo{year}{2018}).

\bibitem[{\citenamefont{Augui\'e and Barnes}(2008)}]{AB08}
\bibinfo{author}{\bibfnamefont{B.}~\bibnamefont{Augui\'e}} \bibnamefont{and}
  \bibinfo{author}{\bibfnamefont{W.~L.} \bibnamefont{Barnes}},
  \bibinfo{journal}{Phys.\ Rev.\ Lett.} \textbf{\bibinfo{volume}{101}},
  \bibinfo{pages}{143902} (\bibinfo{year}{2008}).

\bibitem[{\citenamefont{Vecchi et~al.}(2009)\citenamefont{Vecchi, Giannini, and
  {G\'omez Rivas}}}]{VGG09-2}
\bibinfo{author}{\bibfnamefont{G.}~\bibnamefont{Vecchi}},
  \bibinfo{author}{\bibfnamefont{V.}~\bibnamefont{Giannini}}, \bibnamefont{and}
  \bibinfo{author}{\bibfnamefont{J.}~\bibnamefont{{G\'omez Rivas}}},
  \bibinfo{journal}{Phys.\ Rev.\ Lett.} \textbf{\bibinfo{volume}{102}},
  \bibinfo{pages}{146807} (\bibinfo{year}{2009}).

\bibitem[{\citenamefont{{Augui\'{e}} et~al.}(2010)\citenamefont{{Augui\'{e}},
  {Benda\~{n}a}, Barnes, and {Garc\'{\i}a de Abajo}}}]{paper164}
\bibinfo{author}{\bibfnamefont{B.}~\bibnamefont{{Augui\'{e}}}},
  \bibinfo{author}{\bibfnamefont{X.~M.} \bibnamefont{{Benda\~{n}a}}},
  \bibinfo{author}{\bibfnamefont{W.~L.} \bibnamefont{Barnes}},
  \bibnamefont{and} \bibinfo{author}{\bibfnamefont{F.~J.}
  \bibnamefont{{Garc\'{\i}a de Abajo}}}, \bibinfo{journal}{Phys.\ Rev.\ B}
  \textbf{\bibinfo{volume}{82}}, \bibinfo{pages}{109039}
  (\bibinfo{year}{2010}).

\bibitem[{\citenamefont{Humphrey and Barnes}(2014)}]{HB14}
\bibinfo{author}{\bibfnamefont{A.~D.} \bibnamefont{Humphrey}} \bibnamefont{and}
  \bibinfo{author}{\bibfnamefont{W.~L.} \bibnamefont{Barnes}},
  \bibinfo{journal}{Phys.\ Rev.\ B} \textbf{\bibinfo{volume}{90}},
  \bibinfo{pages}{075404} (\bibinfo{year}{2014}).

\bibitem[{\citenamefont{Humphrey and Barnes}(2016)}]{HB16}
\bibinfo{author}{\bibfnamefont{A.~D.} \bibnamefont{Humphrey}} \bibnamefont{and}
  \bibinfo{author}{\bibfnamefont{W.~L.} \bibnamefont{Barnes}},
  \bibinfo{journal}{J.\ Opt.} \textbf{\bibinfo{volume}{18}},
  \bibinfo{pages}{035005} (\bibinfo{year}{2016}).

\bibitem[{\citenamefont{Kwadrin and Koenderink}(2014)}]{KK14}
\bibinfo{author}{\bibfnamefont{A.}~\bibnamefont{Kwadrin}} \bibnamefont{and}
  \bibinfo{author}{\bibfnamefont{A.~F.} \bibnamefont{Koenderink}},
  \bibinfo{journal}{Phys.\ Rev.\ B} \textbf{\bibinfo{volume}{89}},
  \bibinfo{pages}{045120} (\bibinfo{year}{2014}).

\bibitem[{\citenamefont{Baur et~al.}(2018)\citenamefont{Baur, Sanders, and
  Manjavacas}}]{ama59}
\bibinfo{author}{\bibfnamefont{S.}~\bibnamefont{Baur}},
  \bibinfo{author}{\bibfnamefont{S.}~\bibnamefont{Sanders}}, \bibnamefont{and}
  \bibinfo{author}{\bibfnamefont{A.}~\bibnamefont{Manjavacas}},
  \bibinfo{journal}{ACS Nano} \textbf{\bibinfo{volume}{12}},
  \bibinfo{pages}{1618} (\bibinfo{year}{2018}).

\bibitem[{\citenamefont{Kravets et~al.}(2018)\citenamefont{Kravets, Kabashin,
  Barnes, and Grigorenko}}]{KKB18}
\bibinfo{author}{\bibfnamefont{V.~G.} \bibnamefont{Kravets}},
  \bibinfo{author}{\bibfnamefont{A.~V.} \bibnamefont{Kabashin}},
  \bibinfo{author}{\bibfnamefont{W.~L.} \bibnamefont{Barnes}},
  \bibnamefont{and} \bibinfo{author}{\bibfnamefont{A.~N.}
  \bibnamefont{Grigorenko}}, \bibinfo{journal}{Chem.\ Rev.}
  \textbf{\bibinfo{volume}{118}}, \bibinfo{pages}{5912} (\bibinfo{year}{2018}).

\bibitem[{\citenamefont{Kravets et~al.}(2008)\citenamefont{Kravets, Schedin,
  and Grigorenko}}]{KSG08}
\bibinfo{author}{\bibfnamefont{V.~G.} \bibnamefont{Kravets}},
  \bibinfo{author}{\bibfnamefont{F.}~\bibnamefont{Schedin}}, \bibnamefont{and}
  \bibinfo{author}{\bibfnamefont{A.~N.} \bibnamefont{Grigorenko}},
  \bibinfo{journal}{Phys.\ Rev.\ Lett.} \textbf{\bibinfo{volume}{101}},
  \bibinfo{pages}{087403} (\bibinfo{year}{2008}).

\bibitem[{\citenamefont{Giannini et~al.}(2010)\citenamefont{Giannini, Vecchi,
  and G\'omez~Rivas}}]{GVG10}
\bibinfo{author}{\bibfnamefont{V.}~\bibnamefont{Giannini}},
  \bibinfo{author}{\bibfnamefont{G.}~\bibnamefont{Vecchi}}, \bibnamefont{and}
  \bibinfo{author}{\bibfnamefont{J.}~\bibnamefont{G\'omez~Rivas}},
  \bibinfo{journal}{Phys.\ Rev.\ Lett.} \textbf{\bibinfo{volume}{105}},
  \bibinfo{pages}{266801} (\bibinfo{year}{2010}).

\bibitem[{\citenamefont{Rodriguez
  et~al.}(2012{\natexlab{a}})\citenamefont{Rodriguez, Lozano, Verschuuren,
  Gomes, Lambert, Geyter, Hassinen, Thourhout, Hens, and Rivas}}]{RLV12}
\bibinfo{author}{\bibfnamefont{S.~R.~K.} \bibnamefont{Rodriguez}},
  \bibinfo{author}{\bibfnamefont{G.}~\bibnamefont{Lozano}},
  \bibinfo{author}{\bibfnamefont{M.~A.} \bibnamefont{Verschuuren}},
  \bibinfo{author}{\bibfnamefont{R.}~\bibnamefont{Gomes}},
  \bibinfo{author}{\bibfnamefont{K.}~\bibnamefont{Lambert}},
  \bibinfo{author}{\bibfnamefont{B.~D.} \bibnamefont{Geyter}},
  \bibinfo{author}{\bibfnamefont{A.}~\bibnamefont{Hassinen}},
  \bibinfo{author}{\bibfnamefont{D.~V.} \bibnamefont{Thourhout}},
  \bibinfo{author}{\bibfnamefont{Z.}~\bibnamefont{Hens}}, \bibnamefont{and}
  \bibinfo{author}{\bibfnamefont{J.~G.} \bibnamefont{Rivas}},
  \bibinfo{journal}{Appl.\ Phys.\ Lett.} \textbf{\bibinfo{volume}{100}},
  \bibinfo{pages}{111103} (\bibinfo{year}{2012}{\natexlab{a}}).

\bibitem[{\citenamefont{Schokker and Koenderink}(2014)}]{SK14_2}
\bibinfo{author}{\bibfnamefont{A.~H.} \bibnamefont{Schokker}} \bibnamefont{and}
  \bibinfo{author}{\bibfnamefont{A.~F.} \bibnamefont{Koenderink}},
  \bibinfo{journal}{Phys.\ Rev.\ B} \textbf{\bibinfo{volume}{90}},
  \bibinfo{pages}{155452} (\bibinfo{year}{2014}).

\bibitem[{\citenamefont{Smirnova and Kivshar}(2014)}]{SK14}
\bibinfo{author}{\bibfnamefont{D.~A.} \bibnamefont{Smirnova}} \bibnamefont{and}
  \bibinfo{author}{\bibfnamefont{Y.~S.} \bibnamefont{Kivshar}},
  \bibinfo{journal}{Phys.\ Rev.\ B} \textbf{\bibinfo{volume}{90}},
  \bibinfo{pages}{165433} (\bibinfo{year}{2014}).

\bibitem[{\citenamefont{Lozano et~al.}(2014)\citenamefont{Lozano, Grzela,
  Verschuuren, Ramezani, and Rivas}}]{LGV14}
\bibinfo{author}{\bibfnamefont{G.}~\bibnamefont{Lozano}},
  \bibinfo{author}{\bibfnamefont{G.}~\bibnamefont{Grzela}},
  \bibinfo{author}{\bibfnamefont{M.~A.} \bibnamefont{Verschuuren}},
  \bibinfo{author}{\bibfnamefont{M.}~\bibnamefont{Ramezani}}, \bibnamefont{and}
  \bibinfo{author}{\bibfnamefont{J.~G.} \bibnamefont{Rivas}},
  \bibinfo{journal}{Nanoscale} \textbf{\bibinfo{volume}{6}},
  \bibinfo{pages}{9223} (\bibinfo{year}{2014}).

\bibitem[{\citenamefont{Cotrufo et~al.}(2016)\citenamefont{Cotrufo, Osorio, and
  Koenderink}}]{COK16}
\bibinfo{author}{\bibfnamefont{M.}~\bibnamefont{Cotrufo}},
  \bibinfo{author}{\bibfnamefont{C.~I.} \bibnamefont{Osorio}},
  \bibnamefont{and} \bibinfo{author}{\bibfnamefont{A.~F.}
  \bibnamefont{Koenderink}}, \bibinfo{journal}{ACS Nano}
  \textbf{\bibinfo{volume}{10}}, \bibinfo{pages}{3389} (\bibinfo{year}{2016}).

\bibitem[{\citenamefont{Olson et~al.}(2016)\citenamefont{Olson, Manjavacas,
  Basu, Huang, Schlather, Zheng, Halas, Nordlander, and Link}}]{ama41}
\bibinfo{author}{\bibfnamefont{J.}~\bibnamefont{Olson}},
  \bibinfo{author}{\bibfnamefont{A.}~\bibnamefont{Manjavacas}},
  \bibinfo{author}{\bibfnamefont{T.}~\bibnamefont{Basu}},
  \bibinfo{author}{\bibfnamefont{D.}~\bibnamefont{Huang}},
  \bibinfo{author}{\bibfnamefont{A.~E.} \bibnamefont{Schlather}},
  \bibinfo{author}{\bibfnamefont{B.}~\bibnamefont{Zheng}},
  \bibinfo{author}{\bibfnamefont{N.~J.} \bibnamefont{Halas}},
  \bibinfo{author}{\bibfnamefont{P.}~\bibnamefont{Nordlander}},
  \bibnamefont{and} \bibinfo{author}{\bibfnamefont{S.}~\bibnamefont{Link}},
  \bibinfo{journal}{ACS\ Nano} \textbf{\bibinfo{volume}{10}},
  \bibinfo{pages}{1108} (\bibinfo{year}{2016}).

\bibitem[{\citenamefont{Sung et~al.}(2008)\citenamefont{Sung, Hicks, Van~Duyne,
  and Spears}}]{SHV08}
\bibinfo{author}{\bibfnamefont{J.}~\bibnamefont{Sung}},
  \bibinfo{author}{\bibfnamefont{E.~M.} \bibnamefont{Hicks}},
  \bibinfo{author}{\bibfnamefont{R.~P.} \bibnamefont{Van~Duyne}},
  \bibnamefont{and} \bibinfo{author}{\bibfnamefont{K.~G.}
  \bibnamefont{Spears}}, \bibinfo{journal}{J.\ Phys.\ Chem.\ C}
  \textbf{\bibinfo{volume}{112}}, \bibinfo{pages}{4091} (\bibinfo{year}{2008}).

\bibitem[{\citenamefont{Fedotov et~al.}(2010)\citenamefont{Fedotov,
  Papasimakis, Plum, Bitzer, Walther, Kuo, Tsai, and Zheludev}}]{FPP10}
\bibinfo{author}{\bibfnamefont{V.~A.} \bibnamefont{Fedotov}},
  \bibinfo{author}{\bibfnamefont{N.}~\bibnamefont{Papasimakis}},
  \bibinfo{author}{\bibfnamefont{E.}~\bibnamefont{Plum}},
  \bibinfo{author}{\bibfnamefont{A.}~\bibnamefont{Bitzer}},
  \bibinfo{author}{\bibfnamefont{M.}~\bibnamefont{Walther}},
  \bibinfo{author}{\bibfnamefont{P.}~\bibnamefont{Kuo}},
  \bibinfo{author}{\bibfnamefont{D.~P.} \bibnamefont{Tsai}}, \bibnamefont{and}
  \bibinfo{author}{\bibfnamefont{N.~I.} \bibnamefont{Zheludev}},
  \bibinfo{journal}{Phys.\ Rev.\ Lett.} \textbf{\bibinfo{volume}{104}},
  \bibinfo{pages}{223901} (\bibinfo{year}{2010}).

\bibitem[{\citenamefont{Natarov et~al.}(2011)\citenamefont{Natarov, Byelobrov,
  Sauleau, Benson, and Nosich}}]{NBS11}
\bibinfo{author}{\bibfnamefont{D.~M.} \bibnamefont{Natarov}},
  \bibinfo{author}{\bibfnamefont{V.~O.} \bibnamefont{Byelobrov}},
  \bibinfo{author}{\bibfnamefont{R.}~\bibnamefont{Sauleau}},
  \bibinfo{author}{\bibfnamefont{T.~M.} \bibnamefont{Benson}},
  \bibnamefont{and} \bibinfo{author}{\bibfnamefont{A.~I.}
  \bibnamefont{Nosich}}, \bibinfo{journal}{Opt.\ Express}
  \textbf{\bibinfo{volume}{19}}, \bibinfo{pages}{22176} (\bibinfo{year}{2011}).

\bibitem[{\citenamefont{Rodriguez
  et~al.}(2012{\natexlab{b}})\citenamefont{Rodriguez, Schaafsma, Berrier, and
  G\'omez-Rivas}}]{RSB12}
\bibinfo{author}{\bibfnamefont{S.~R.~K.} \bibnamefont{Rodriguez}},
  \bibinfo{author}{\bibfnamefont{M.~C.} \bibnamefont{Schaafsma}},
  \bibinfo{author}{\bibfnamefont{A.}~\bibnamefont{Berrier}}, \bibnamefont{and}
  \bibinfo{author}{\bibfnamefont{J.}~\bibnamefont{G\'omez-Rivas}},
  \bibinfo{journal}{Physica\ B} \textbf{\bibinfo{volume}{407}},
  \bibinfo{pages}{4081} (\bibinfo{year}{2012}{\natexlab{b}}).

\bibitem[{\citenamefont{Matsushima}(2017)}]{M17}
\bibinfo{author}{\bibfnamefont{A.}~\bibnamefont{Matsushima}},
  \bibinfo{journal}{Proc.\ of 2017 IEEE International Conference on
  Computational Electromagnetics (ICCEM)} pp. \bibinfo{pages}{236--237}
  (\bibinfo{year}{2017}).

\bibitem[{\citenamefont{Martikainen et~al.}(2017)\citenamefont{Martikainen,
  Moilanen, and T{\"o}rm{\"a}}}]{MMT17}
\bibinfo{author}{\bibfnamefont{J.-P.} \bibnamefont{Martikainen}},
  \bibinfo{author}{\bibfnamefont{A.~J.} \bibnamefont{Moilanen}},
  \bibnamefont{and}
  \bibinfo{author}{\bibfnamefont{P.}~\bibnamefont{T{\"o}rm{\"a}}},
  \bibinfo{journal}{Philos.\ Trans.\ Royal Soc.\ A}
  \textbf{\bibinfo{volume}{375}} (\bibinfo{year}{2017}).

\bibitem[{\citenamefont{Zhao et~al.}(2003)\citenamefont{Zhao, Kelly, and
  Schatz}}]{ZKS03}
\bibinfo{author}{\bibfnamefont{L.}~\bibnamefont{Zhao}},
  \bibinfo{author}{\bibfnamefont{K.~L.} \bibnamefont{Kelly}}, \bibnamefont{and}
  \bibinfo{author}{\bibfnamefont{G.~C.} \bibnamefont{Schatz}},
  \bibinfo{journal}{J.\ Phys.\ Chem.\ B} \textbf{\bibinfo{volume}{107}},
  \bibinfo{pages}{7343} (\bibinfo{year}{2003}).

\bibitem[{\citenamefont{Steshenko and Capolino}(2009)}]{C09}
\bibinfo{author}{\bibfnamefont{S.}~\bibnamefont{Steshenko}} \bibnamefont{and}
  \bibinfo{author}{\bibfnamefont{F.}~\bibnamefont{Capolino}},
  \emph{\bibinfo{title}{Theory and phenomena of metamaterials, Chapter 8}}
  (\bibinfo{publisher}{CRC press}, \bibinfo{year}{2009}).

\bibitem[{\citenamefont{Teperik and Degiron}(2012)}]{TD12}
\bibinfo{author}{\bibfnamefont{T.~V.} \bibnamefont{Teperik}} \bibnamefont{and}
  \bibinfo{author}{\bibfnamefont{A.}~\bibnamefont{Degiron}},
  \bibinfo{journal}{Phys.\ Rev.\ B} \textbf{\bibinfo{volume}{86}},
  \bibinfo{pages}{245425} (\bibinfo{year}{2012}).

\bibitem[{\citenamefont{Myroshnychenko
  et~al.}(2008)\citenamefont{Myroshnychenko, {Rodr\'{\i}guez-Fern\'andez},
  Pastoriza-Santos, Funston, Novo, Mulvaney, {Liz-Marz\'an}, and {Garc\'{\i}a
  de Abajo}}}]{paper112}
\bibinfo{author}{\bibfnamefont{V.}~\bibnamefont{Myroshnychenko}},
  \bibinfo{author}{\bibfnamefont{J.}~\bibnamefont{{Rodr\'{\i}guez-Fern\'andez}}},
  \bibinfo{author}{\bibfnamefont{I.}~\bibnamefont{Pastoriza-Santos}},
  \bibinfo{author}{\bibfnamefont{A.~M.} \bibnamefont{Funston}},
  \bibinfo{author}{\bibfnamefont{C.}~\bibnamefont{Novo}},
  \bibinfo{author}{\bibfnamefont{P.}~\bibnamefont{Mulvaney}},
  \bibinfo{author}{\bibfnamefont{L.~M.} \bibnamefont{{Liz-Marz\'an}}},
  \bibnamefont{and} \bibinfo{author}{\bibfnamefont{F.~J.}
  \bibnamefont{{Garc\'{\i}a de Abajo}}}, \bibinfo{journal}{Chem.\ Soc.\ Rev.}
  \textbf{\bibinfo{volume}{37}}, \bibinfo{pages}{1792} (\bibinfo{year}{2008}).

\bibitem[{\citenamefont{Johnson and Christy}(1972)}]{JC1972}
\bibinfo{author}{\bibfnamefont{P.~B.} \bibnamefont{Johnson}} \bibnamefont{and}
  \bibinfo{author}{\bibfnamefont{R.~W.} \bibnamefont{Christy}},
  \bibinfo{journal}{Phys.\ Rev.\ B} \textbf{\bibinfo{volume}{6}},
  \bibinfo{pages}{4370} (\bibinfo{year}{1972}).

\bibitem[{\citenamefont{Koppens et~al.}(2011)\citenamefont{Koppens, Chang, and
  {Garc\'{\i}a de Abajo}}}]{paper176}
\bibinfo{author}{\bibfnamefont{F.~H.~L.} \bibnamefont{Koppens}},
  \bibinfo{author}{\bibfnamefont{D.~E.} \bibnamefont{Chang}}, \bibnamefont{and}
  \bibinfo{author}{\bibfnamefont{F.~J.} \bibnamefont{{Garc\'{\i}a de Abajo}}},
  \bibinfo{journal}{Nano\ Lett.} \textbf{\bibinfo{volume}{11}},
  \bibinfo{pages}{3370} (\bibinfo{year}{2011}).

\bibitem[{\citenamefont{Chen et~al.}(2012)\citenamefont{Chen, Badioli,
  Alonso-Gonz\'alez, Thongrattanasiri, Huth, Osmond, Spasenovi\'c, Centeno,
  Pesquera, Godignon et~al.}}]{paper196}
\bibinfo{author}{\bibfnamefont{J.}~\bibnamefont{Chen}},
  \bibinfo{author}{\bibfnamefont{M.}~\bibnamefont{Badioli}},
  \bibinfo{author}{\bibfnamefont{P.}~\bibnamefont{Alonso-Gonz\'alez}},
  \bibinfo{author}{\bibfnamefont{S.}~\bibnamefont{Thongrattanasiri}},
  \bibinfo{author}{\bibfnamefont{F.}~\bibnamefont{Huth}},
  \bibinfo{author}{\bibfnamefont{J.}~\bibnamefont{Osmond}},
  \bibinfo{author}{\bibfnamefont{M.}~\bibnamefont{Spasenovi\'c}},
  \bibinfo{author}{\bibfnamefont{A.}~\bibnamefont{Centeno}},
  \bibinfo{author}{\bibfnamefont{A.}~\bibnamefont{Pesquera}},
  \bibinfo{author}{\bibfnamefont{P.}~\bibnamefont{Godignon}},
  \bibnamefont{et~al.}, \bibinfo{journal}{Nature}
  \textbf{\bibinfo{volume}{487}}, \bibinfo{pages}{77} (\bibinfo{year}{2012}).

\bibitem[{\citenamefont{Yan et~al.}(2012)\citenamefont{Yan, Li, Chandra,
  Tulevski, Wu, Freitag, Zhu, Avouris, and Xia}}]{YLC12}
\bibinfo{author}{\bibfnamefont{H.}~\bibnamefont{Yan}},
  \bibinfo{author}{\bibfnamefont{X.}~\bibnamefont{Li}},
  \bibinfo{author}{\bibfnamefont{B.}~\bibnamefont{Chandra}},
  \bibinfo{author}{\bibfnamefont{G.}~\bibnamefont{Tulevski}},
  \bibinfo{author}{\bibfnamefont{Y.}~\bibnamefont{Wu}},
  \bibinfo{author}{\bibfnamefont{M.}~\bibnamefont{Freitag}},
  \bibinfo{author}{\bibfnamefont{W.}~\bibnamefont{Zhu}},
  \bibinfo{author}{\bibfnamefont{P.}~\bibnamefont{Avouris}}, \bibnamefont{and}
  \bibinfo{author}{\bibfnamefont{F.}~\bibnamefont{Xia}},
  \bibinfo{journal}{Nat.\ Nanotechnol.} \textbf{\bibinfo{volume}{7}},
  \bibinfo{pages}{330} (\bibinfo{year}{2012}).

\bibitem[{\citenamefont{Thongrattanasiri
  et~al.}(2012)\citenamefont{Thongrattanasiri, Koppens, and {Garc\'{\i}a de
  Abajo}}}]{paper182}
\bibinfo{author}{\bibfnamefont{S.}~\bibnamefont{Thongrattanasiri}},
  \bibinfo{author}{\bibfnamefont{F.~H.~L.} \bibnamefont{Koppens}},
  \bibnamefont{and} \bibinfo{author}{\bibfnamefont{F.~J.}
  \bibnamefont{{Garc\'{\i}a de Abajo}}}, \bibinfo{journal}{Phys.\ Rev.\ Lett.}
  \textbf{\bibinfo{volume}{108}}, \bibinfo{pages}{047401}
  (\bibinfo{year}{2012}).

\bibitem[{\citenamefont{Fang et~al.}(2014)\citenamefont{Fang, Wang, Schlather,
  Liu, Ajayan, {Garc\'{\i}a de Abajo}, Nordlander, Zhu, and Halas}}]{paper230}
\bibinfo{author}{\bibfnamefont{Z.}~\bibnamefont{Fang}},
  \bibinfo{author}{\bibfnamefont{Y.}~\bibnamefont{Wang}},
  \bibinfo{author}{\bibfnamefont{A.}~\bibnamefont{Schlather}},
  \bibinfo{author}{\bibfnamefont{Z.}~\bibnamefont{Liu}},
  \bibinfo{author}{\bibfnamefont{P.~M.} \bibnamefont{Ajayan}},
  \bibinfo{author}{\bibfnamefont{F.~J.} \bibnamefont{{Garc\'{\i}a de Abajo}}},
  \bibinfo{author}{\bibfnamefont{P.}~\bibnamefont{Nordlander}},
  \bibinfo{author}{\bibfnamefont{X.}~\bibnamefont{Zhu}}, \bibnamefont{and}
  \bibinfo{author}{\bibfnamefont{N.~J.} \bibnamefont{Halas}},
  \bibinfo{journal}{Nano\ Lett.} \textbf{\bibinfo{volume}{14}},
  \bibinfo{pages}{299} (\bibinfo{year}{2014}).

\bibitem[{\citenamefont{Rodrigo et~al.}(2015)\citenamefont{Rodrigo, Limaj,
  Janner, Etezadi, {Garc\'{\i}a de Abajo}, Pruneri, and Altug}}]{paper256}
\bibinfo{author}{\bibfnamefont{D.}~\bibnamefont{Rodrigo}},
  \bibinfo{author}{\bibfnamefont{O.}~\bibnamefont{Limaj}},
  \bibinfo{author}{\bibfnamefont{D.}~\bibnamefont{Janner}},
  \bibinfo{author}{\bibfnamefont{D.}~\bibnamefont{Etezadi}},
  \bibinfo{author}{\bibfnamefont{F.~J.} \bibnamefont{{Garc\'{\i}a de Abajo}}},
  \bibinfo{author}{\bibfnamefont{V.}~\bibnamefont{Pruneri}}, \bibnamefont{and}
  \bibinfo{author}{\bibfnamefont{H.}~\bibnamefont{Altug}},
  \bibinfo{journal}{Science} \textbf{\bibinfo{volume}{349}},
  \bibinfo{pages}{165} (\bibinfo{year}{2015}).

\bibitem[{\citenamefont{Zundel and Manjavacas}(2017)}]{ama53}
\bibinfo{author}{\bibfnamefont{L.}~\bibnamefont{Zundel}} \bibnamefont{and}
  \bibinfo{author}{\bibfnamefont{A.}~\bibnamefont{Manjavacas}},
  \bibinfo{journal}{ACS Photonics} \textbf{\bibinfo{volume}{4}},
  \bibinfo{pages}{1831} (\bibinfo{year}{2017}).

\bibitem[{\citenamefont{{Garc\'{\i}a de Abajo}}(2014)}]{paper235}
\bibinfo{author}{\bibfnamefont{F.~J.} \bibnamefont{{Garc\'{\i}a de Abajo}}},
  \bibinfo{journal}{ACS\ Photon.} \textbf{\bibinfo{volume}{1}},
  \bibinfo{pages}{135} (\bibinfo{year}{2014}).

\bibitem[{\citenamefont{Yu et~al.}(2017)\citenamefont{Yu, Cox, Saavedra, and
  {Garc\'{\i}a de Abajo}}}]{paper303}
\bibinfo{author}{\bibfnamefont{R.}~\bibnamefont{Yu}},
  \bibinfo{author}{\bibfnamefont{J.~D.} \bibnamefont{Cox}},
  \bibinfo{author}{\bibfnamefont{J.~R.~M.} \bibnamefont{Saavedra}},
  \bibnamefont{and} \bibinfo{author}{\bibfnamefont{F.~J.}
  \bibnamefont{{Garc\'{\i}a de Abajo}}}, \bibinfo{journal}{ACS Photonics}
  \textbf{\bibinfo{volume}{4}}, \bibinfo{pages}{3106} (\bibinfo{year}{2017}).

\bibitem[{\citenamefont{Jenkins and Ruostekoski}(2012)}]{JR12}
\bibinfo{author}{\bibfnamefont{S.~D.} \bibnamefont{Jenkins}} \bibnamefont{and}
  \bibinfo{author}{\bibfnamefont{J.}~\bibnamefont{Ruostekoski}},
  \bibinfo{journal}{Phys.\ Rev.\ B} \textbf{\bibinfo{volume}{86}},
  \bibinfo{pages}{205128} (\bibinfo{year}{2012}).

\bibitem[{\citenamefont{Papasimakis et~al.}(2009)\citenamefont{Papasimakis,
  Fedotov, Fu, Tsai, and Zheludev}}]{PFF09}
\bibinfo{author}{\bibfnamefont{N.}~\bibnamefont{Papasimakis}},
  \bibinfo{author}{\bibfnamefont{V.~A.} \bibnamefont{Fedotov}},
  \bibinfo{author}{\bibfnamefont{Y.~H.} \bibnamefont{Fu}},
  \bibinfo{author}{\bibfnamefont{D.~P.} \bibnamefont{Tsai}}, \bibnamefont{and}
  \bibinfo{author}{\bibfnamefont{N.~I.} \bibnamefont{Zheludev}},
  \bibinfo{journal}{Phys.\ Rev.\ B} \textbf{\bibinfo{volume}{80}},
  \bibinfo{pages}{041102} (\bibinfo{year}{2009}).

\bibitem[{\citenamefont{Augui\'e and Barnes}(2009)}]{AB09}
\bibinfo{author}{\bibfnamefont{B.}~\bibnamefont{Augui\'e}} \bibnamefont{and}
  \bibinfo{author}{\bibfnamefont{W.~L.} \bibnamefont{Barnes}},
  \bibinfo{journal}{Opt.\ Lett.} \textbf{\bibinfo{volume}{34}},
  \bibinfo{pages}{401} (\bibinfo{year}{2009}).

\bibitem[{\citenamefont{Schokker and Koenderink}(2015)}]{SK15}
\bibinfo{author}{\bibfnamefont{A.~H.} \bibnamefont{Schokker}} \bibnamefont{and}
  \bibinfo{author}{\bibfnamefont{A.~F.} \bibnamefont{Koenderink}},
  \bibinfo{journal}{ACS Photonics} \textbf{\bibinfo{volume}{2}},
  \bibinfo{pages}{1289} (\bibinfo{year}{2015}).

\end{thebibliography}

\end{document}